\documentclass[preprint,12pt,authoryear]{elsarticle}
\usepackage{enumitem}
\usepackage[utf8]{inputenc}

\usepackage{lineno}

\usepackage{hyperref}
\usepackage{amssymb}
\usepackage{amsmath}
\usepackage{hyperref}
\usepackage{subcaption}
\usepackage{graphicx}
\usepackage{multicol}
\usepackage{multirow}
\usepackage{booktabs}
\usepackage{lmodern}

\journal{Advances in Space Research}
\begin{document}

\begin{frontmatter}



\title{Assessment of sky brightness in central Portugal:  the Dark Sky destination "Aldeias do Xisto" case study}

\author[inst1]{Domingos Barbosa}
\author[inst2]{Dalmiro Maia}
\author[inst2]{Bruno Morgado} 
\author[inst3]{Pedro Cruz} 
\author[inst2]{Bernardo Relvas}
\author[inst4]{Valério Ribeiro} 
\author[inst5,inst6]{Allan Kardec de Almeida Jr.} 
\author[inst6]{Timothée Vaillant} 
            
\address[inst1]{High Performance Computing Chair, University of Évora, R. Romão Ramalho 59, Évora, 7000, Portugal}

\address[inst2]{Observatório Astronómico Manuel de Barros, Faculdade de Ciências da Universidade do Porto, Alameda do Monte da Virgem, Vila Nova de Gaia, 4430-146, Portugal}

\address[inst3]{Departamento de Física e Astronomia, Faculdade de Ciências da Universidade do Porto, Rua do Campo Alegre, 4169-007 Porto, Portugal}

\address[inst4]{KLM, Data \& Technology - Platform Ground, PO Box 7700, 1117 ZL Schiphol, The Netherlands}

\address[inst5]{Faculty of Sciences and Technology, University of the Azores, Rua da Mãe de Deus, 9500-321 Ponta Delgada, Portugal}

\address[inst6]{CFisUC, Departamento de Física, Universidade de Coimbra, 3004-516 Coimbra, Portugal}



\begin{abstract}
A combination of remote sensing techniques, open space data and field testing was developed to derive astronomical and meteorological parameters and assess Dark Sky brightness of the hitherto unstudied Schist Villages region in the mountainous Interior Center of Portugal. This study covered the period 2012-2017 and established the methodology for regular light pollution and sky brightness studies in the region using satellite imagery to reduce cost and time and increase accuracy of the final analysis. Long-term Artificial Light At Night (ALAN) data from VIIRS instrument aboard the Suomi-NPP satellite along with cloud cover inferred from MODUS/AQUA satellite data and orographic data, were analyzed in a Geographical Information Systems (GIS) environment. The derived ALAN values were consistent with a preliminary site survey selection, confirming the accuracy of the identified dark sky spots in the region. The analyzed region area spans 1624 Km$^2$ and the darkest site within it showed a sky brightness of 21.4 mag/arcsec$^2$. Ground-based SQM photometer observations of the Night Sky Brightness (NSB) at the zenith validated the satellite derived ALAN data. The devised field testing is itself a low cost solution that can be used by teams in developing countries to easily assess dark sky potential prior to the development of local observatories and astro-tourism activities. This study supported the initial certification of this region as a Dark Sky Starlight Tourist Destination since 2018, an important milestone for dark sky protection in central Portugal.
\end{abstract}



\begin{keyword}
 Dark sky  \sep artificial light at night \sep Site selection \sep astro-tourism \sep Geographic information system \sep Remote sensing
\end{keyword}

\end{frontmatter}


\section{Introduction}

The decade of the 2020s may be remembered by history, in part, as an era of two distinct but directly connected phenomena: a time of transformation in the use of orbital space in the vicinity of the Earth, and the impact that industrial transformation had on the appearance of the night sky \citep{2022NatAs...6.1120B,2023arXiv231102184B}. 

Nowadays, most of the population in industrialized nations live under skies where the artificial light at night (ALAN) exceeds natural light by thousands of times. The luminous flux produced by artificial lighting is sent directly or indirectly towards the sky raising the sky background flux, a problem exacerbated in and near urban areas. The main effect of ALAN is the loss of perception of the Universe around us, as the increase in the brightness of the night sky due to artificial light prevents the vision of the starry sky. Besides the loss of a starry dark sky, ALAN presents a strong environmental impact since it negatively affects the biosphere, flora and fauna, and human life \citep{Hermoso2022,yak2022,Ferretti2025}.
Additionally, astronomical or space science observatories are usually installed at dark sky territories for long term science service. Those research infrastructures' potential for cutting edge science and capacity building may be strongly impacted since the concurring ALAN and space satellite constellations development trends may risk decades of billions of euros of global long term investments.

As an early demonstration of the importance of dark skies for astronomy, we note that Carl Friedrich Gauss (1777-1855) derived a method for orbit determination, where he predicted the position of the dwarf planet Ceres in October 1801, based on the data obtained by the astronomer Giuseppe Piazzi (1746-1826) with a telescope in Palermo with only 3 degrees apart for the considered observation time interval in January 1st of this same year. This first real solution to the Lambert Problem \citep{jordan1964application} could not be tested without an open and clean access to the observation of the Sky, allowing the discovery of three new other planets subsequently and entering into a new era of celestial mechanics and astronomy discoveries.  

\subsection{Dark Sky preservation initiatives}

The night Sky is now recognized as natural resource with a cultural and historic dimension that closely intertwines humankind with the nocturnal environment. Several organizations certify places with appropriate outdoor lighting policies and practices that minimize sky glow, that are compatible with nature conservation, and that offer interpretation and outreach programs to residents and visitors alike. As of November 2025, and according to the The Dark Skies Advisory Group (DSAG) at the International Union for Conservation of Nature (IUCN), there are over over 325 dark sky and starlight tourism locations across at least 34 countries globally, from strictly protected areas to astronomical research sites, from biosphere reserves to towns and villages\footnote{{\url https://darkskyparks.org/dark-skies-and-nature-conservation/}}\citep{Dusi2026,Iwanicki2026}. Indeed, the IUCN recognizes the importance of natural darkness to nature conservation, to ecological integrity of protected areas, and to the sustainability of environment for healthier cities \citep{IUCN2024}. Furthermore, this has been recognized by its economic impact on astro-tourism prompting even global hotel chains to adopt light pollution mitigation through the development of corporate strategies and regulatory approaches \citep{Iwanicki2026}.

More recently, concerns raised on ALAN produced by the exponential growth of space satellite megaconstellations led the Committee on Space Research (COSPAR) to promote a Dark and Quiet Skies Initiative with the United Nations Office for Outer Space Affairs (UNOOSA) and the International Astronomical Union (IAU) \citep{author2026_cospar_abstract,copuos2026_crp35,cospar2026press}. Under this initiative that brought astronomy, satellite industries, policy, and law makers, it is recognized the importance of the preservation of optical astronomy from artificial light ("dark") and radio astronomy from frequency interference ("quiet"). This position echoes the IAU Centre for the Protection of the Dark and Quiet Sky from Satellite Constellation Interference (IAU CPS) manifesto for the protection of the dark sky \cite{IAUCPS}.


The International Dark-Sky Association (IDA) has championed the protection of the night skies for present and future generations with pioneering activities against light pollution \citep{IDSA2011}, promoting mutually beneficial solutions that allow people to appreciate dark, star-filled skies while enjoying the benefits of responsible development aspects. 
 UNESCO, the United Nations World Tourism Organization (UNWTO) and the International Astronomical Union (IAU) jointly recognized the benefits of protection of Dark Skies and developed guidelines and promoted certifications through the Dark Sky Starlight Foundation at Instituto de Astrofísica de Canarias for Tourism Destinations. These certifications are issued to protect and defend the sky and valuing it as a resource necessary for life and the intangible heritage of humanity \citep{2011IAUS..260..449M,Dalgleish2021} after the ‘‘La Palma Declaration’’ that further emphasizes the protection and conservation of the night skies as an important scientific, cultural, environmental and tourist resource \citep{2011IAUS..260..449M}. 

 Geographic Information Systems (GIS), coupled to MultiCriteria Decision Analysis (MCDA)-based methodology using space data are used as important tools for site suitability assessment of astronomy Observatories from Europe to Middle East \citep{Duarte2016,Yılmaz2023,2023CoSka..53a..28Y,Aladwan2026}. This methodology combines environmental and atmospheric variables, including cloud cover, digital elevation models, and in-situ artificial light pollution measurements. These can be quickly integrated in site suitability identification on a global scale and incorporate hybrid weighting approaches or robustness evaluation through sensitivity analysis \citep{Duarte2016,Aladwan2026}. Application of these techniques has been extended to the investigation of potential regions with strong potential for astro-tourism, including dark sky parks, reserves, and other recent astro-tourism initiatives, like in the Balcans and Southwestern T\"urkiye regions \citep{Kurt2024,Aksaker2025}.

\section{Site selection and Methods}

For ground-based astronomy the brightness of the night sky is the limiting factor that determines the efficiency of any particular telescope in terms of detecting faint objects \citep{barentine2023}. 
Light scattered in the lower atmosphere, known as ‘skyglow’, reduces the contrast of astronomical objects against the background night sky and impairs their visibility \citep{2022NatAs...6.1120B}, while light emitted at small negative elevation angles can reach far from its source into sensitive natural spaces \citep{2016MNRAS.458..438K}. This has known negative consequences for living organisms including humans and nowadays, it is recognized that more than 80\% of the Earth population is exposed to light pollution and at least 30\% cannot see anymore the Milky Way with the naked eye \citep{2016SciAdv...2e0377F}. This  represents energy waste typically accompanied by an increase in ground-level illumination at significant distances from human settlements \citep{Luginbuhl2013} and imposes growing limits on the study of the cosmos (\citep{Green2022}). 

Site testing campaigns and analysis of satellite data are usually performed within the classical scheme of optical seeing properties, air transparency, sky darkness, and cloudiness. Starlight Dark Sky Tourist Destination certification requires compliance with a minimum set of criteria:
\begin{enumerate}
    \item Seeing better than 3 arcsec.
    \item A cloud free sky.
    \item Visual magnitude of the sky higher than 21 mag/arcsec$^2$.
    \item Sky transparency such that the Naked Eye Limiting Magnitude (NELM) in the visible is higher than 6.
    \item All of these parameters above must be met simultaneously in at least 50\% of the nights per year.
\end{enumerate}

\subsection{The CIM-Coimbra region}

Although the study leveraged the Starlight astro-tourism designation certification, we aimed to identify sites with a strong potential for astro-tourism, rather than focus solely on a Dark Sky international certification. The analysis targeted a data timeseries between 2012 and 2017 and analyzed the territory of CIM-Coimbra, a Portuguese community established since 2013 that aggregates several smaller municipalities spanning a large and very diverse territory from coastal areas to the central mountains, covering an area of $\sim 4336$ km$^2$. We plan a second second study paper to analyze long term time series until 2025 and enable extraction of long term trends in light pollution and changes in the regional sky background that may be useful to guide future dark sky protection policies for the benefit of future generations.

The CIM-Coimbra coastal plains are densely populated but a double mountain chain about 100 km away from the coast geographically shields the very sparsely populated CIM-Coimbra mountain region from the industrialized and light polluted coast. Interestingly, several areas of the mountain territory within this area are classified for their biodiversity and environmental interest as Natura 2000, a network of protected areas covering Europe's most valuable and threatened species and habitats \citep{Popescu2014,Hermoso2022}. 

\begin{figure}
\centerline{
\includegraphics[width=.85\textwidth]{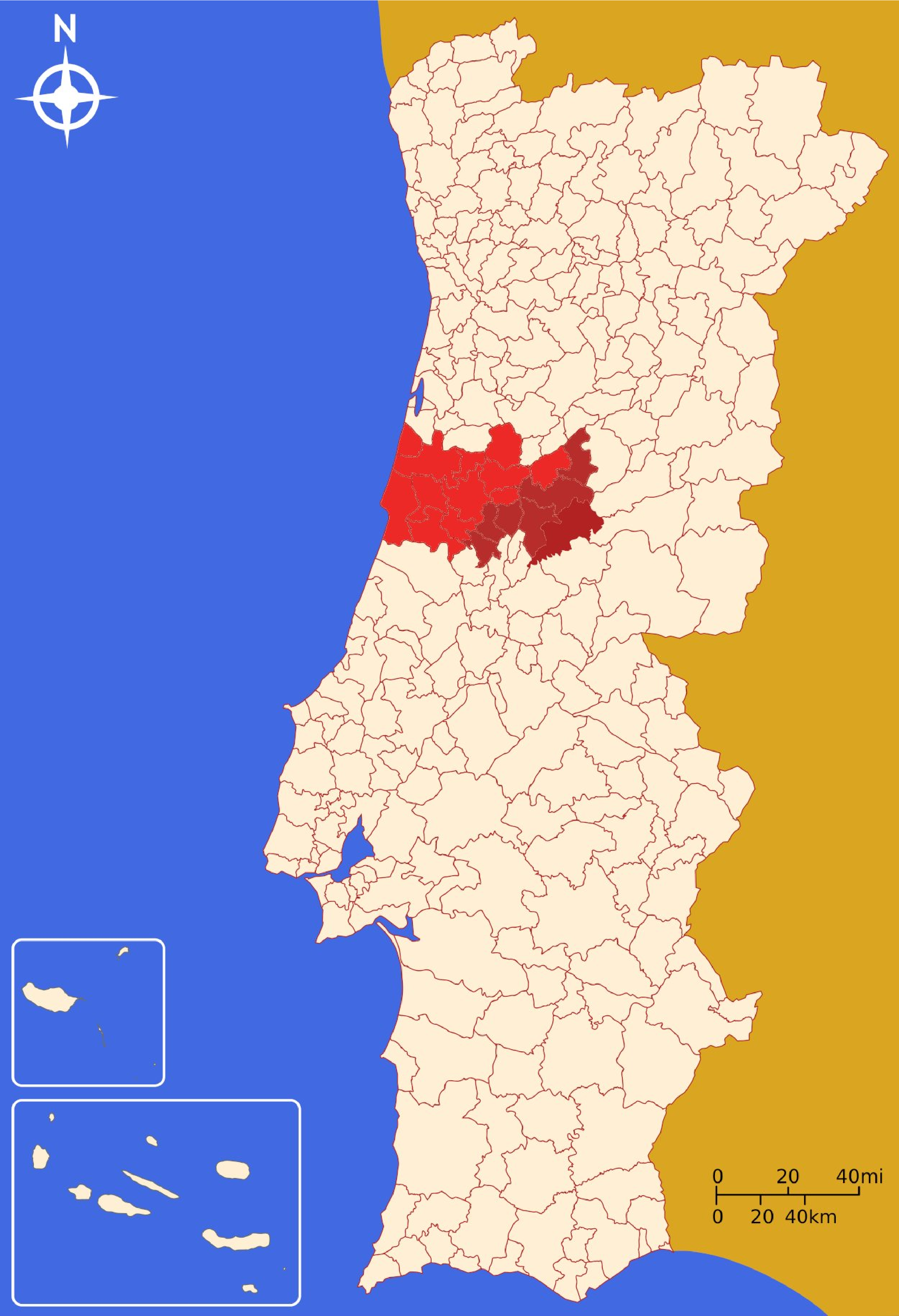}
}
\caption[Map of all municipalities in Portugal]{Map of all municipalities in Portugal. CIM-Coimbra municipalities: highlighted in red; the region in dark red:  selected region for application of the developed methodology using Earth observations' data and GIS tools.}
\label{fg:setecim}
\end{figure}

There are 21 intermunicipal communities in Portugal, and one of their competences is promoting local development, social cohesion and attract new avenues for impact activities like tourism. CIM-Coimbra intermunicipal community integrates 19 municipalities.
Figure~\ref{fg:setecim} shows an administrative map of Portugal and the studied region.

The area is dark red (Figure~\ref{fg:setecim}) was already known for its dark and quiet sky and was chosen since 2006 after an initial site selection to install a first radiotelescope for the Galactic Emission Mapping project in Portugal (GEM-P) \citep{2006NewA...11..551F}. This led to the early development of the Pampilhosa da Serra Space Observatory (PASO), an observatory thought as the astronomical epicentre of a new dark Sky destination \citep{barbosa2024paso}. Launched as an active observatory in expansion, PASO astronomical capabilities and advanced Space Domain Awareness (SDA) capabilities fed research groups spread across the country and enabled cutting-edge development of astrodynamics applications to surveillance of the space around Earth \citep{akajtfcnelder} and the cislunar space \citep{akajtangential}, the research  of low-thrust propulsion systems (such as a solar sail) for mission design for asteroid rendez-vous \citep{almeida24solar}, analysis of mitigation methods of space debris. PASO was supported by the ENGAGE SKA Research Infrastructure, and part of local capacitation contributions to the Square Kilometer Array Observatory \citep{skao}. Connected to these efforts, the recently started project SPACE \footnote{https://compete2030.gov.pt/operacoes/listagem/COMPETE2030-FEDER-00860300/} develops studies in orbit determination of satellites based on the early request of observational optical and radar data from the sensors at PASO region. PASO provided support to local outreach activities and citizen science campaigns such as \citep{2022AcAau.200..612B}, an early example of the continuous application of digital telescopes in a pedagogical, citizen-science and pandemic context. More recently and following the award of Dark Sky Starlight Tourism Destination, an open dome platform for Astronomical observation (Geoscope\footnote{https://starlight.aldeiasdoxisto.pt/pt/geoscope/}) was installed for astro-tourism and astronomical outreach at Faj\~ao, one of the dark sky locations considered in this study.

A first selection within the main town territories in CIM-Coimbra region was made based on the results of the semi-empirical models presented in \citep{2001MNRAS.328..689C}. These maps include the Portuguese mainland (their figures 2 and 3) and from them we can infer that only a smaller part (seven out of the nineteen municipalities) of CIM-Coimbra region are likely to fulfill the last two certification criteria. This smaller inter-municipality area of 1624 km$^2$ and the focus of this work, is highlighted by a darker shade of red in Figure 1.1, and encompasses  Arganil, Góis, Lousã, Miranda do Corvo, Oliveira do Hospital, Pampilhosa da Serra, and Penela municipalities. ( Figures \ref{fg:aod} and \ref{fg:nuvenspt} show the contours and coordinates of this area).

 
Within this region, the Towns of Schist ({\em Rede das Aldeias do Xisto}) network represent 27 local traditional settlements spread over the mountainous Interior Center in Portugal. They are organized in four subsets: (i) twelve towns in the Lousã mountain range group, (ii) five towns in the Açor mountain range group, six towns in the Zêzere river group, and (iv) four towns in the Tejo-Ocreza rivers group. For the case study analysis we will only consider the 23 towns in the first three groups; Table~\ref{tb:xisto} indicates names, geographic coordinates and orography of those settlements.

\begin{table}
\caption[Names and coordinates of Towns from Schist]{Names and coordinates of the 23 Towns from Schist within the vicinity of the dark red region in Figure~\ref{fg:setecim}.}
\label{tb:xisto}
\begin{center}
\begin{tabular}{@{} *2l *3c @{}}
\toprule
 \multicolumn{1}{c}{Group} & \multicolumn{1}{c}{Town}    &  Latitude & Longitude & Altitude \\
   &  &   &  &   (meters)  \\\midrule
\multirow{12}{*}{Lousã Mountains} & \href{https://aldeiasdoxisto.pt/sites/default/files/Mapa%20Aigra%20Nova.pdf}{Aigra Nova} & 40\textdegree  07’12”N & 8\textdegree 09’15”W & 615 \\
& \href{https://aldeiasdoxisto.pt/sites/default/files/Mapa%20Aigra%20Velha.pdf}{Aigra Velha} & 40\textdegree  06’31”N & 8\textdegree 08’57”W & 770\\
& \href{https://aldeiasdoxisto.pt/sites/default/files/Mapa%20Candal.pdf}{Candal} & 40\textdegree  04’50”N & 8\textdegree 12’12”W & 630\\
& \href{https://aldeiasdoxisto.pt/sites/default/files/Mapa%20Casal%20de%20S.%20Sima%CC%83o.pdf}{Casal de São Simão} & 39\textdegree  55’01”N & 8\textdegree 19’21”W & 305\\
& \href{https://aldeiasdoxisto.pt/sites/default/files/Mapa%20Casal%20Novo.pdf}{Casal Novo} & 40\textdegree  05’29”N & 8\textdegree 14’09”W & 550 \\
& \href{https://aldeiasdoxisto.pt/sites/default/files/planta_cerdeira.pdf}{Cerdeira} & 40\textdegree  05’38”N & 8\textdegree 11’41”W & 660\\
& \href{https://aldeiasdoxisto.pt/sites/default/files/Mapa%20Chiqueiro.pdf}{Chiqueiro} & 40\textdegree  05’10”N & 8\textdegree 13’57”W & 650\\
& \href{https://aldeiasdoxisto.pt/sites/default/files/Mapal%20Comareira.pdf}{Comareira} & 40\textdegree  07’37”N & 8\textdegree 09’11”W & 515\\
& \href{https://aldeiasdoxisto.pt/sites/default/files/Mapa%20Ferraria%20Sa%CC%83o%20Joa%CC%83o.pdf}{Ferraria de São João} & 39\textdegree  58’26”N & 8\textdegree 19’27”W & 640\\
& \href{https://aldeiasdoxisto.pt/sites/default/files/Mapa%20Gondramaz.pdf}{Gondramaz} & 40\textdegree  03’43”N & 8\textdegree 16’22”W & 640 \\
& \href{https://aldeiasdoxisto.pt/sites/default/files/Mapa%20Pena.pdf}{Pena} &40\textdegree  06’37”N & 8\textdegree 08’06”W & 590\\& \href{https://aldeiasdoxisto.pt/sites/default/files/Mapa%20Talasnal.pdf}{Talasnal} & 40\textdegree  05’29”N & 8\textdegree 13’32”W & 500\\[1ex]

\multirow{5}{*}{Açor Mountains}& \href{https://aldeiasdoxisto.pt/sites/default/files/Mapa%20de%20Aldeia%20das%20Dez.pdf}{Aldeia das Dez} & 40\textdegree  17’40”N & 7\textdegree 51’56”W & 460\\
& \href{https://aldeiasdoxisto.pt/sites/default/files/Mapa%20da%20Aldeia%20da%20Benfeita.pdf}{Benfeita} & 40\textdegree  13’44”N & 7\textdegree 56’41”W & 300 \\
& \href{https://aldeiasdoxisto.pt/sites/default/files/Mapa%20da%20Aldeia%20da%20Benfeita.pdf}{Fajão} & 40\textdegree  08’56”N & 7\textdegree 55’22”W & 720\\
& \href{https://aldeiasdoxisto.pt/sites/default/files/Mapa%20da%20Aldeia%20de%20Sobral%20de%20S%C3%A3o%20Miguel.pdf}{Sobral de São Miguel} & 40\textdegree  12’40”N & 7\textdegree44’33”W & 560\\
& \href{https://aldeiasdoxisto.pt/sites/default/files/Mapa%20da%20Aldeia%20de%20Vila%20Cova%20de%20Alva.pdf}{Vila Cova de Alva}  & 40\textdegree  16’59”N & 7\textdegree56’31”W & 236 \\
[1ex]

\multirow{6}{*}{Zêzere River} & \href{https://aldeiasdoxisto.pt/sites/default/files/Mapa%20da%20Aldeia%20de%20%C3%81lvaro.pdf}{Álvaro} & 39\textdegree  58’32”N & 7\textdegree 57’57”W & 350 \\
& \href{https://aldeiasdoxisto.pt/sites/default/files/Mapa%20da%20Aldeia%20da%20Barroca.pdf}{Barroca} & 40\textdegree  06’32”N & 7\textdegree 43’03”W & 375\\
& \href{https://aldeiasdoxisto.pt/sites/default/files/Mapa%20da%20Aldeia%20de%20Janeiro%20de%20Baixo.pdf}{Janeiro de Baixo} & 40\textdegree 02’47”N & 7\textdegree 48’18”W & 303 \\
& \href{https://aldeiasdoxisto.pt/sites/default/files/Mapa%20da%20Aldeia%20de%20Janeiro%20de%20Cima.pdf}{Janeiro de Cima} & 40\textdegree  04’00”N & 7\textdegree48’00”W & 340\\
& \href{https://aldeiasdoxisto.pt/sites/default/files/Mapa%20da%20Aldeia%20de%20Mosteiro.pdf}{Mosteiro} & 39\textdegree  56’11”N & 8\textdegree 11’10”W & 301 \\
& \href{https://aldeiasdoxisto.pt/sites/default/files/Mapa%20da%20Aldeia%20de%20Pedrog%C3%A3o%20Pequeno.pdf}{Pedrógão Pequeno} &39\textdegree  54’13”N & 8\textdegree 07’56”W & 350 \\\bottomrule 
 \end{tabular}
\end{center}
\end{table}

\section{Satellite observations}
\label{sec:sat}

Sky brightness is affected by contributions from human made sources of light and by related atmospheric parameters like cloud cover and the atmospheric aerosol content. These parameters play an important role in astronomical site conditions. Although in situ data is important to study these parameters, the studied region has an abrupt mountainous orography and it experiences cyclic forest fires. In particular, unusually severe forest fire conditions were reported in 2017 and a series of strong storms in the first months of 2018. 
Hence, our analysis relied mostly on satellite data provided by the Defense Meteorological Satellite Program (DMSP) operated by the National Oceanic and Atmospheric Administration (NOAA) and its successor Suomi National Polar-orbiting Partnership (Suomi-NPP) satellite. These  satellites operate in synchronous solar low altitude orbits (830 km) with an orbital period of \~101 minutes and provide vital environmental, meteorological, and nighttime imagery data for civilian and military applications. In particular, data obtained by the Visible-Infrared Imager - Radiometer Suite (VIIRS) \citep{liang2014,Choi2022} experiment, one of the key five instruments aboard the Suomi-NPP satellite launched in October 2011 \citep{Goldberg2018}, was extensively used in our sky brightness at night analysis. VIIRS is a scanning radiometer that collects visible and infrared imagery and radiometric measurements of the land, atmosphere, cryosphere, and oceans and its day-night band (DNB) represents a major advancement in nighttime imaging capabilities from space \citep{Lee2006,Miller2012,cao2014,Liao2013}. A VIIRS instrument is also present in NOAA-20 and NOAA-21 satellites and thus these satellites provide a global visible and infrared imagery monitoring of wildfires, smoke, sea ice, and nighttime lights (Day/Night Band) with a 3,000-km swath, offering daily coverage for environmental and weather forecasting.

\section{Sky transparency}
Atmospheric extinction is essentially the absorption, scattering, and reflection of light from cosmic sources by tiny solid and liquid aerosol particles suspended in the atmosphere that decrease the quantity of light that reaches the surface.
The aerosol types that are mostly seen over this Portuguese territory originates from biomass burning aerosols (controlled fires for forest cleaning or uncontrolled summer forest fires) and windblown desert dust from African (Sahara and Sahel) origin. Due to their size, those aerosols can reduce visibility in the optical wavelength range and  can therefore affect astronomical observations. 

Aerosol Atmospheric extinction data of Satellite origin is often defined as Aerosol Optical Depth ($AOD$):
\begin{equation}
    AOD= \log \frac{I_a}{I_g}
\end{equation}

\noindent
where $I_g$ refers to the intensity measured at the ground and $I_a$ to the intensity that would be measured if the Earth's atmosphere was fully transparent.
$AOD$ indicates how much direct light is prevented from reaching the ground by the aerosol particles. $AOD$ is a dimensionless number that is related to the amount of aerosols in the vertical column of atmosphere over the observation location. Expressing extinction in a logarithmic scale as a magnitude, we get:
\begin{equation}
    m_{ext}= 2.5 \log_{\mbox{10}}\frac{I_a}{I_g}
\end{equation}

Extinction in magnitudes is thus 1.086 times the $AOD$. A large $AOD$ indicates elevated levels of absorbing aerosols in the Earth's atmosphere. However, a smaller threshold indicating a clear atmosphere may not indicate suitability for astronomical purposes. Based on an analysis of $AOD$ values measured during dusty episodes at the Canary Islands, \citep{2008MNRAS.391..507V} suggests that $AOD$ in the V band (550 nm) should be lower than 0.1, to separate a clear sky from a dusty sky. The criterion for Starlight Dark Sky Touristic Destination certification is not as demanding: the sky does not need to be fully transparent and some "dustiness" is allowed as long as the Naked Eye Limiting Magnitude (NELM) in the V band remains above 6 magnitudes. 

NELM is a quantity that can vary substantially from observer to observer, but it can be expressed statistically as a function of  $AOD$ or other related extinction coefficients for different kinds of observers. From  \citep{1990PASP..102..212S} a typical site with a sky brightness of 21 mag/arcsec$^2$ in the V band, and for a typical observer, a NELM $>$ 6 is obtained for an atmospheric extinction below 0.343 mag, which corresponds to 0.316 $AOD$. As such we consider that an $AOD$ threshold above 0.3 corresponds to a dusty, non-transparent sky.

\subsection{Analysis of VIIRS aerosol data}
\label{sec:aod}

Since 2013, VIIRS Aerosol Optical Thickness and Aerosol Particle Size Parameter Intermediate Product (IP) and Environmental Data Record (EDR) data were released to the public with a Stage 2 Validation level. This assessment was based on both qualitative and quantitative analysis, and the VIIRS aerosol products were compared with aerosol products obtained from instruments onboard other satellites. Those data are available for download and consultation\footnote{\href{https://www.star.nesdis.noaa.gov/jpss/EDRs/products_aerosols.php}{STAR JPSS EDR website}}.  $AOD$ is retrieved for clear (cloud-free) pixels during daytime observations with zenith angles $\theta \leq 80^\circ$. This means that from those data we can determine the fraction of cloudless days which are non-transparent by determining at each pixel the ratio between the number of measures of $AOD> 0.3$ and the number of measures available.

\begin{figure}
\centerline{
\includegraphics[width=.45\textwidth]{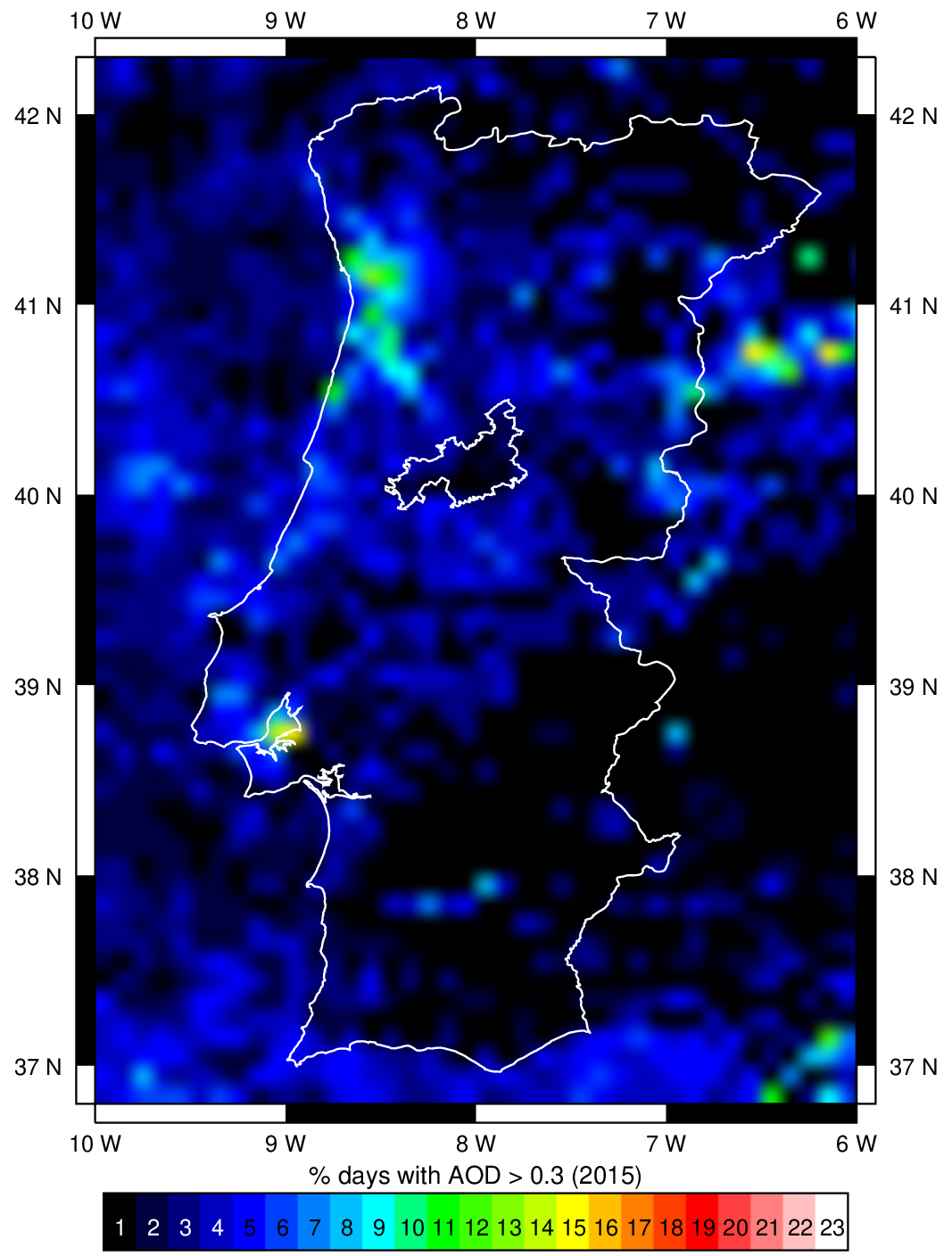}\includegraphics[width=.45\textwidth]{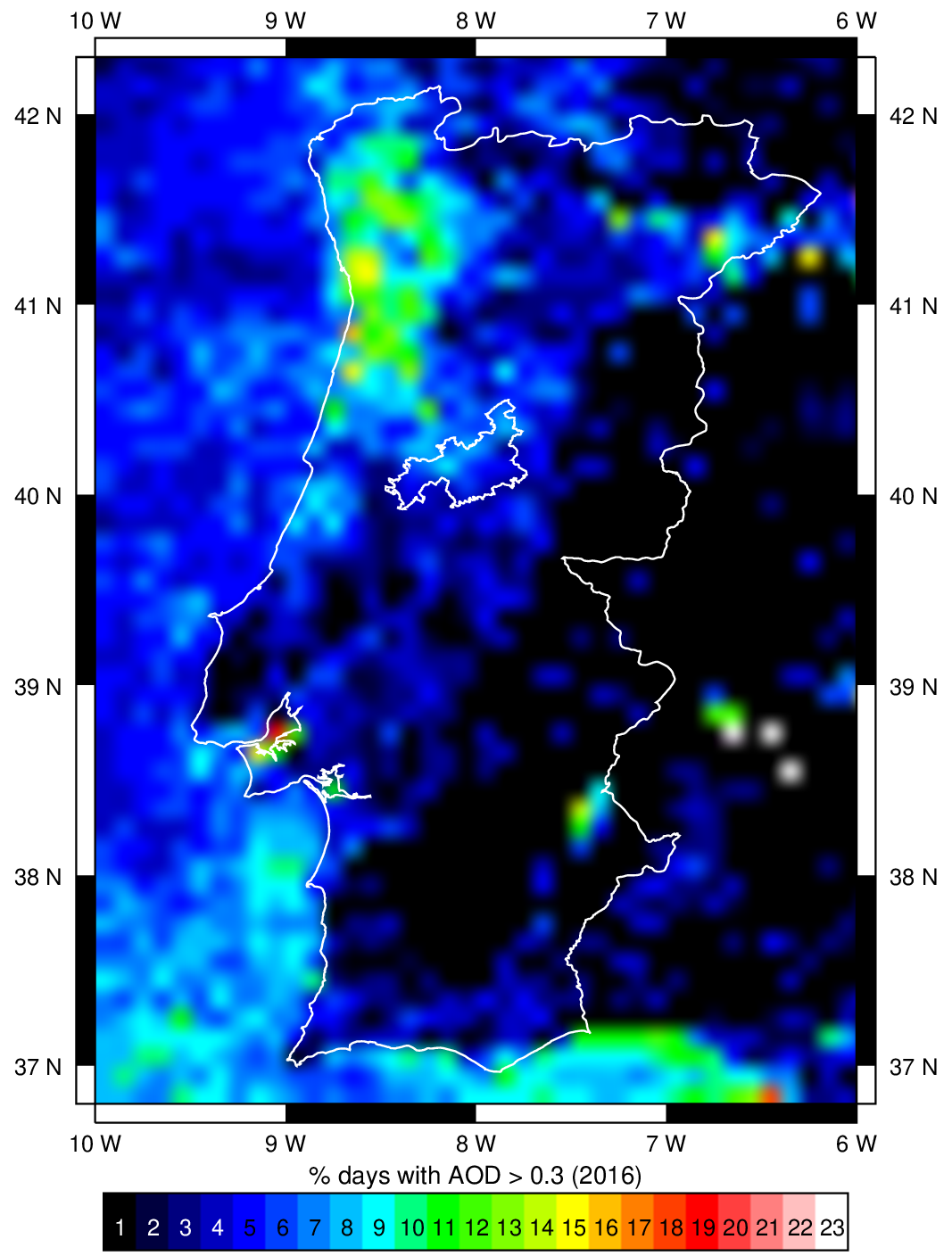}
}
\caption[Transparency maps]{Maps of the fraction of dusty days over the Portuguese mainland for the years 2015 (left) and 2016 (right). The contours of the area defined by the seven councils of reference are superposed on the map.}
\label{fg:aod}
\end{figure}

We present in Figure~\ref{fg:aod} two maps showing the number of days for the years 2015 and 2016 when the $AOD$ measured by VIIRS is above the 0.3 threshold, ie the  maps indicate the fraction of cloudless days that are dusty days. For those two years, and also for 2017, the available $AOD$ data was at 0.1 degree spatial resolution. There is also data available for 2012, 2013 and 2014, but at 0.25 degree spatial resolution. We focus on 2015 and 2016 because those were the only years when high spatial resolution cloud cover could be estimated. We produced maps for the other years and they are consistent with the 2015 and 2016 maps: an average of 9\% of the cloudless days fail the NELM threshold over the seven municipalities of reference.

\section{Nighttime cloud cover}
VIIRS nighttime sensor (for the Day/Night Band, or DNB) provides important nighttime imagery products. Many of these DNB imagery products are intended for use in near real-time applications, such as monitoring, managing, and analyzing on-going natural hazards or events; its primary purpose is supporting the short-term weather prediction and disaster response communities. These real-time products are not intended for use in scientific research, but the NASA Suomi-NPP Land Science Investigator-led Processing System (SIPS) routinely releases nighttime lights products with more extensive processing, that are corrected for clouds, lunar reflectance, and other instrument-related artifacts that may be missed in the DNB near real-time products. These products will be used extensively for the yearly night sky brightness maps described in Section\ref{sec:skybrightness}.

\begin{figure}
\centerline{
\includegraphics[width=.45\textwidth]{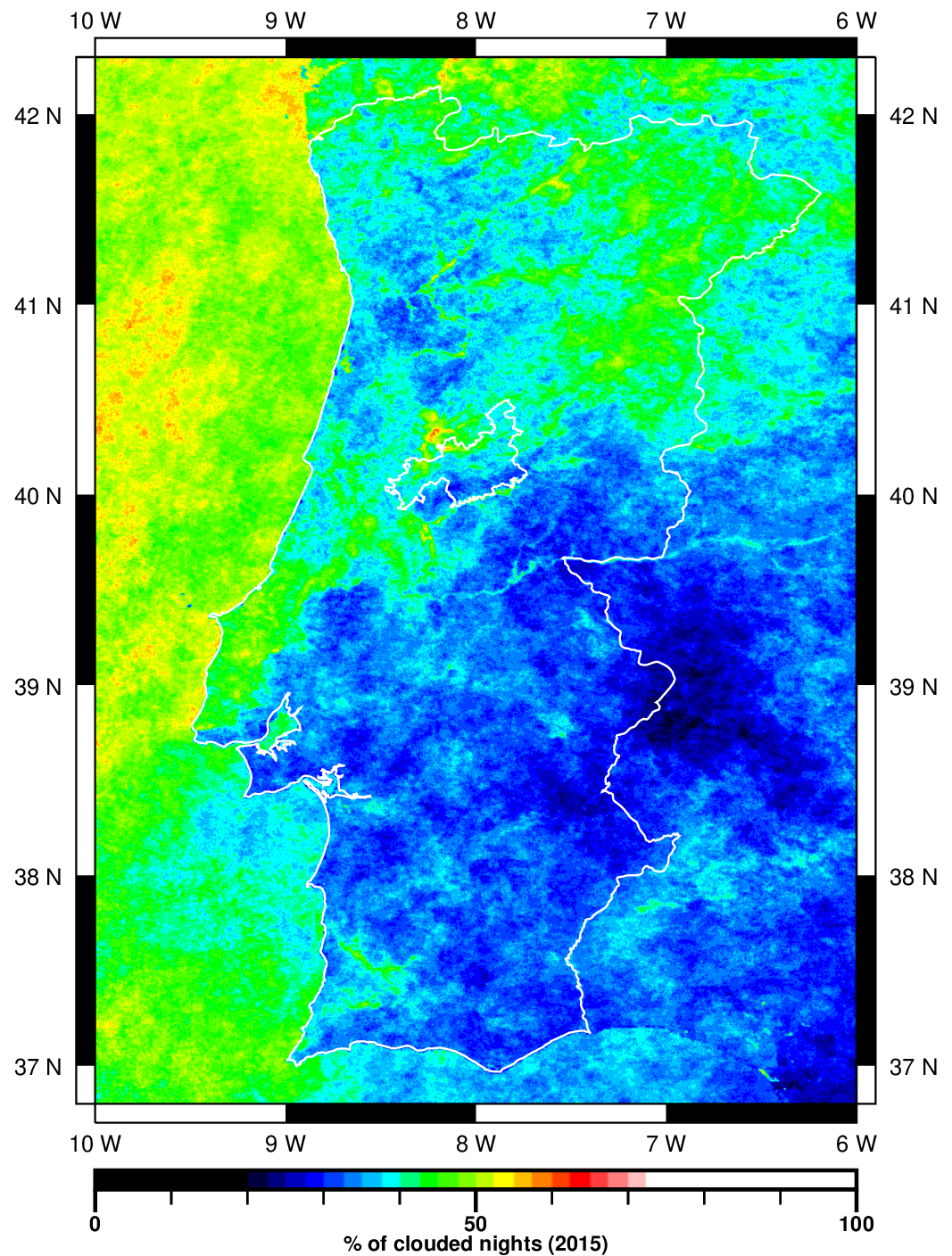}\includegraphics[width=.45\textwidth]{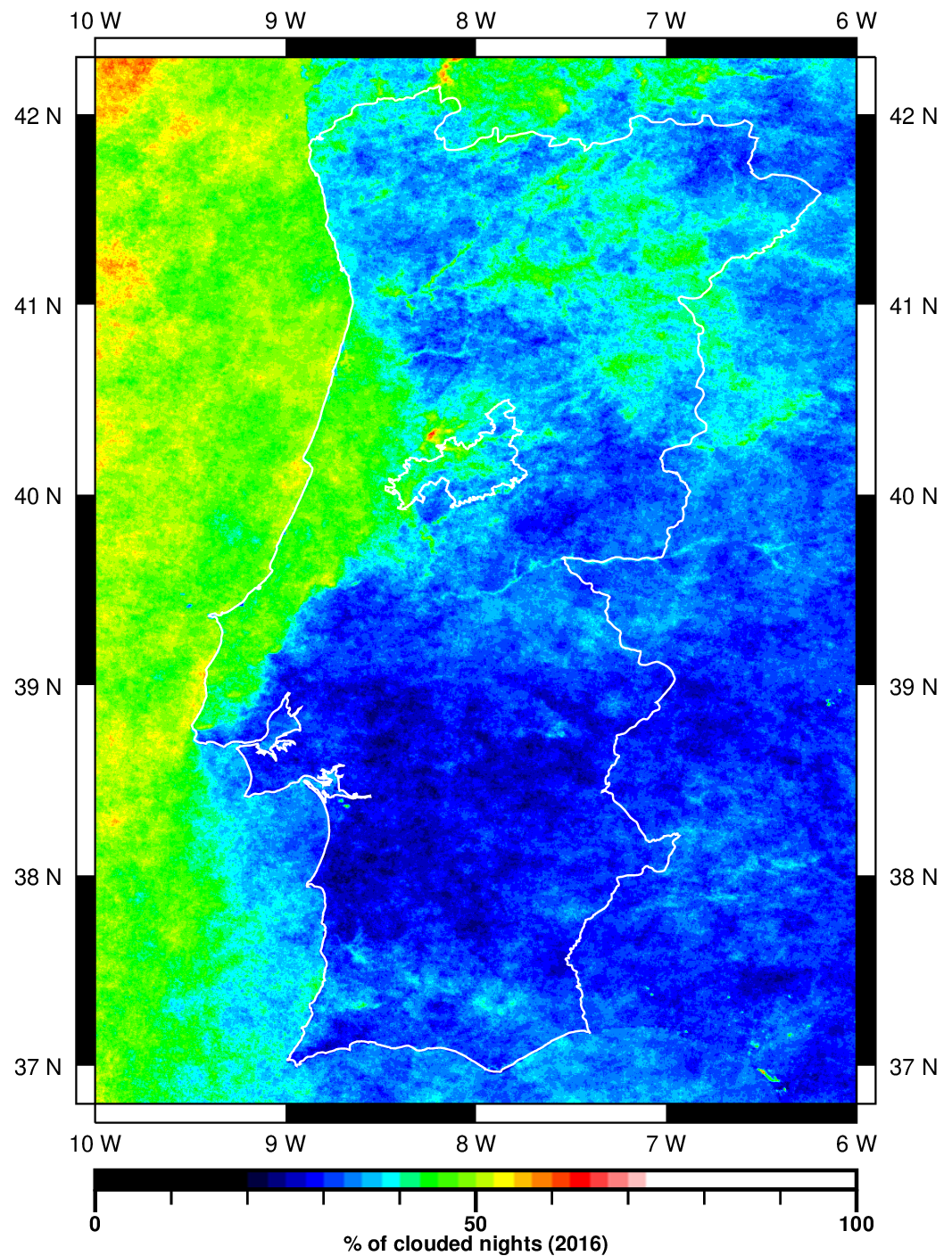}
}
\caption[Clouded nights over Portugal, VIIRS data]{Map of the number of clouded nights over Portugal for the years 2015 (left) and 2016 (right) taken as good and indicative examples of the decade (not very much affected by forest fires and extreme storm patterns). The contours of the area defined by the seven councils of reference are superposed on the map.}
\label{fg:nuvenspt}
\end{figure}

As accessory information to the brightness maps, the DMSP team provides georeferenced maps of the number of nights when observations were made, and the number of those observations that were rejected due to the presence of clouds. Those maps have a resolution of 15 arcsec, corresponding to 500 meters at Equator and 350 meters at the latitudes of our region of interest, and from them we can compute a map of the average  cloud cover during the nighttime period by dividing the number of observations per pixel that were refused due to the presence of clouds by the total number of observations that were recorded for the same pixel. Unfortunately, although VIIRS observations started in the year 2012, at the time of this analysis only the yearly brightness maps and accessory information for the years 2015 and 2016 were readily available. From those data we have computed maps of nighttime cloud cover for the whole Portuguese mainland. Those are shown in Figure~\ref{fg:nuvenspt}, illustrating the fact that our reference councils sit at the boundary between a cloudy littoral to the West and a cloudless interior to the East. 

\begin{figure}
\begin{center}
\includegraphics[width=.85\textwidth]{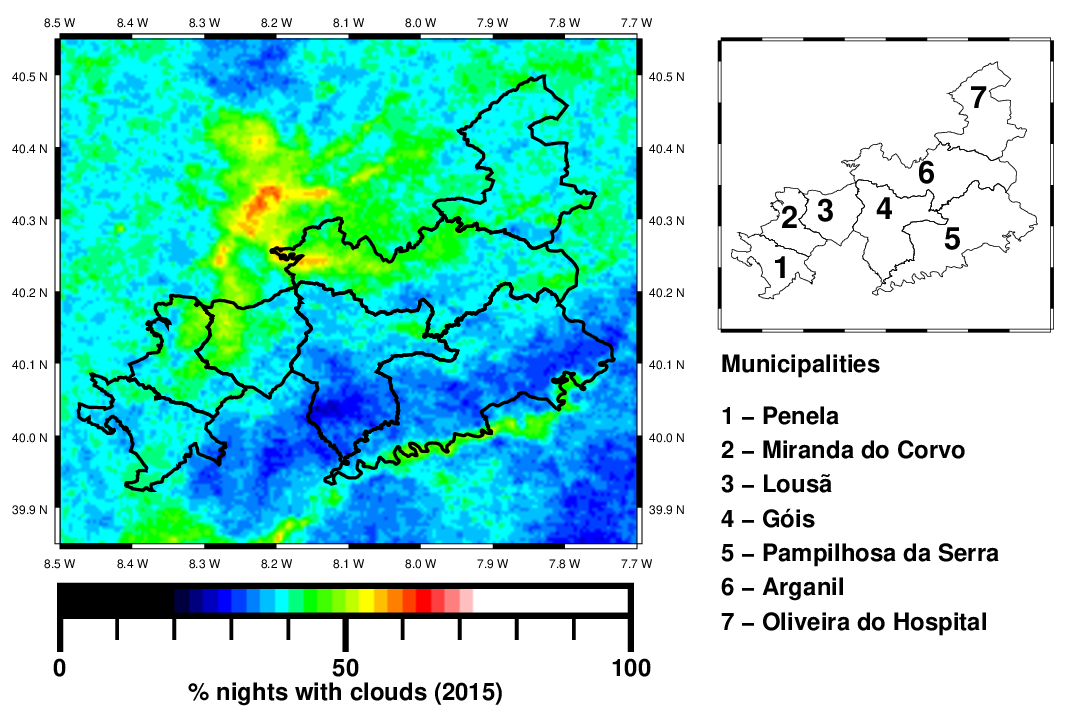}
\end{center}
\caption[Clouded nights for the region of interest in 2015]{Map of the number of clouded nights over the region of interest for the year 2015.}
\label{fg:nuvensa}
\end{figure}

The resolution of the VIIRS maps is such that we can even observe detailed variation over the region of interest. Figure~\ref{fg:nuvensa} shows a zoom over the seven reference councils for the year 2015. The cloudy strip that we see south of Pampilhosa da Serra corresponds to the river Z\^ezere and reflects the foggy conditions that are often encountered on its margins. Figure \ref{fg:nuvensa} shows that over most of the area of interest, for the year 2015, more than 60\% of the nights were cloudless. The exceptions are some parishes in the North-west parts of the councils of Arganil, and Lous\~a, that approach or even exceed the 50\% of nights with clouds.

\begin{figure}
\begin{center}
\includegraphics[width=.85\textwidth]{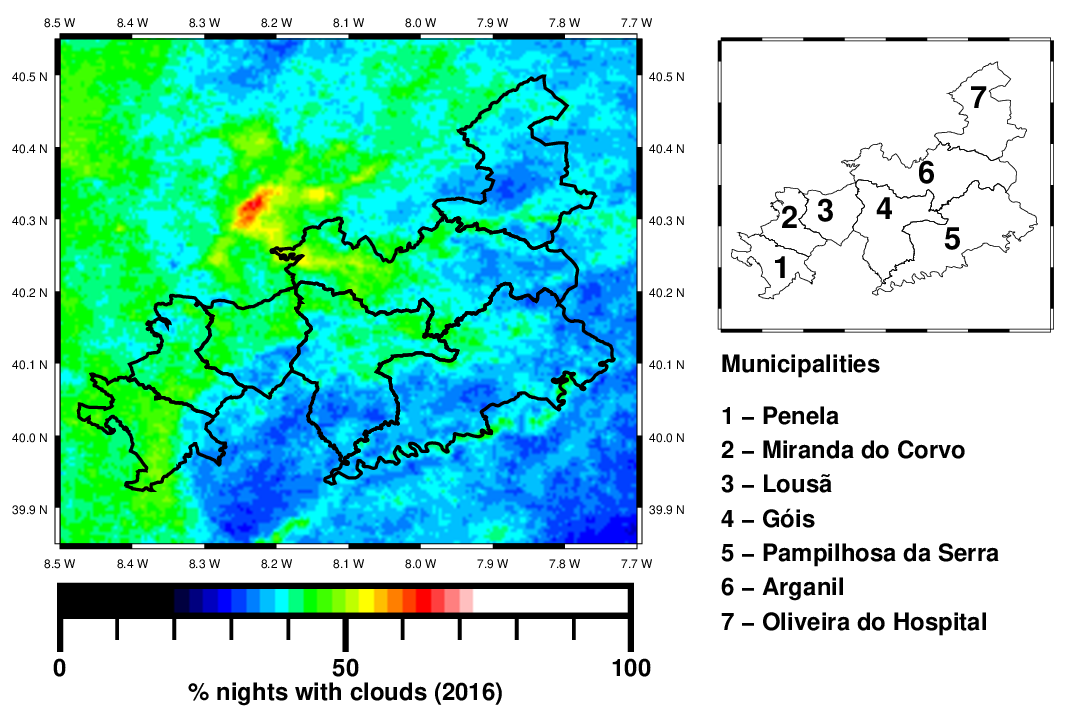}
\end{center}
\caption[Clouded nights for the region of interest in 2016]{Map of the number of clouded nights over the region of interest for the year 2016.}
\label{fg:nuvensb}
\end{figure}

Figure~\ref{fg:nuvensb} shows the situation for the year 2016, and the overall picture is very similar. The year 2016 was a little bit cloudier than the preceding year over the region of interest. A comparison of the cloud cover in Figure~\ref{fg:nuvensb} with the digital elevation model in Figure~\ref{fg:demcim} reveals the night clouds accumulate mostly to the North and West of the Serra do Açor and Serra da Lous\~a mountain ranges (effect of natural barrier to the North and West dominated winds from the coast).

\section{Preliminary region for certification}

\begin{figure}
\begin{center}
\includegraphics[width=.85\textwidth]{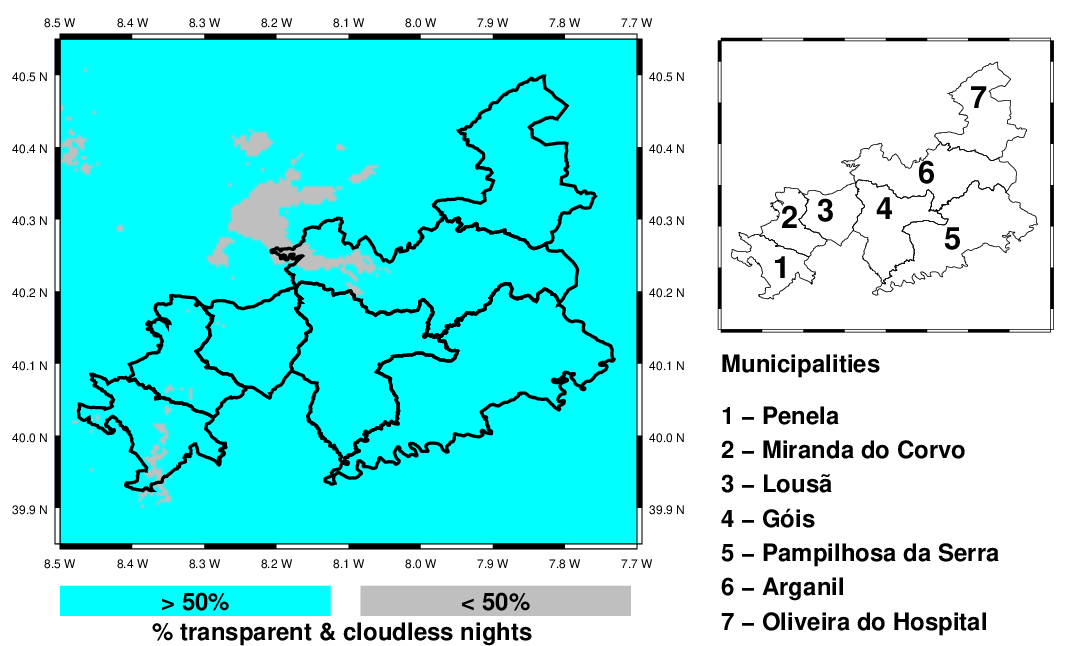}
\end{center}
\caption[Region fulfilling transparency and cloud cover requirements]{Map of the region fulfilling simultaneously the transparency and cloud cover requirements for more than 50\% of the nights (in cyan).}
\label{fg:cloudless}
\end{figure}

For Starlight Tourist Destination certification purposes more than 50\% of the days in the year must be simultaneously cloudless and transparent. As seen in section~\ref{sec:aod} over our region of interest something like 9\% of the cloudless days per year are non-transparent. This means that for a given location to be considered both cloudless and transparent on more than 50\% of the days, it should be cloudless on more than 55\% of the days. Figure~\ref{fg:cloudless} shows the result of applying such a threshold to the cloud cover data obtained by averaging the cloud cover information for the years 2015 and 2016. This criterion does not exclude a significant area, only small portions of Arganil and Penela fail the cloud cover and transparency thresholds.

\subsection{Long term analysis}
We restricted our analysis to 2015 and 2016 since those were the two years when high-resolution cloud cover data were available from VIIRS.
Much longer data series were available from other experiments, in particular the CLOUD FRACTION product provided by the Moderate Resolution Imaging Spectroradiometer (MODIS) aboard the NASA AQUA Earth-Observing Satellite Mission \citep{Platnick2003, Ackerman2017,Platnick2017,Roman2024}. This product is available since 2002, but it has a much lower spatial resolution, roughly 4.2 km at our latitudes and it provides cloud information for the day period only, not for the nights. Nonetheless, although not exactly comparable to VIIRS data, it can be used to check if 2015 and 2016 are typical years regarding cloud cover. 

\begin{figure}
\begin{center}
\includegraphics[width=.3\textwidth]{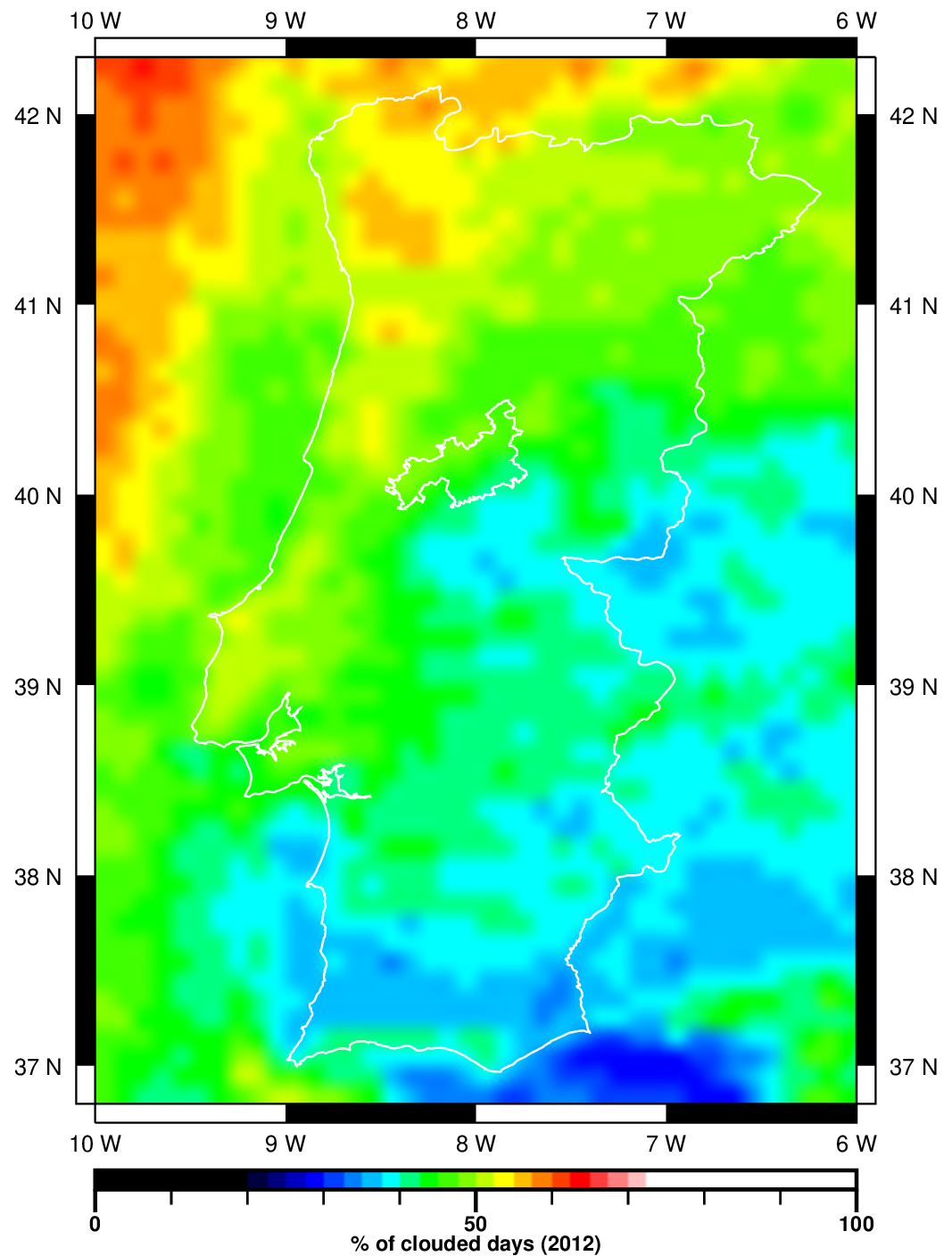}
\includegraphics[width=.3\textwidth]{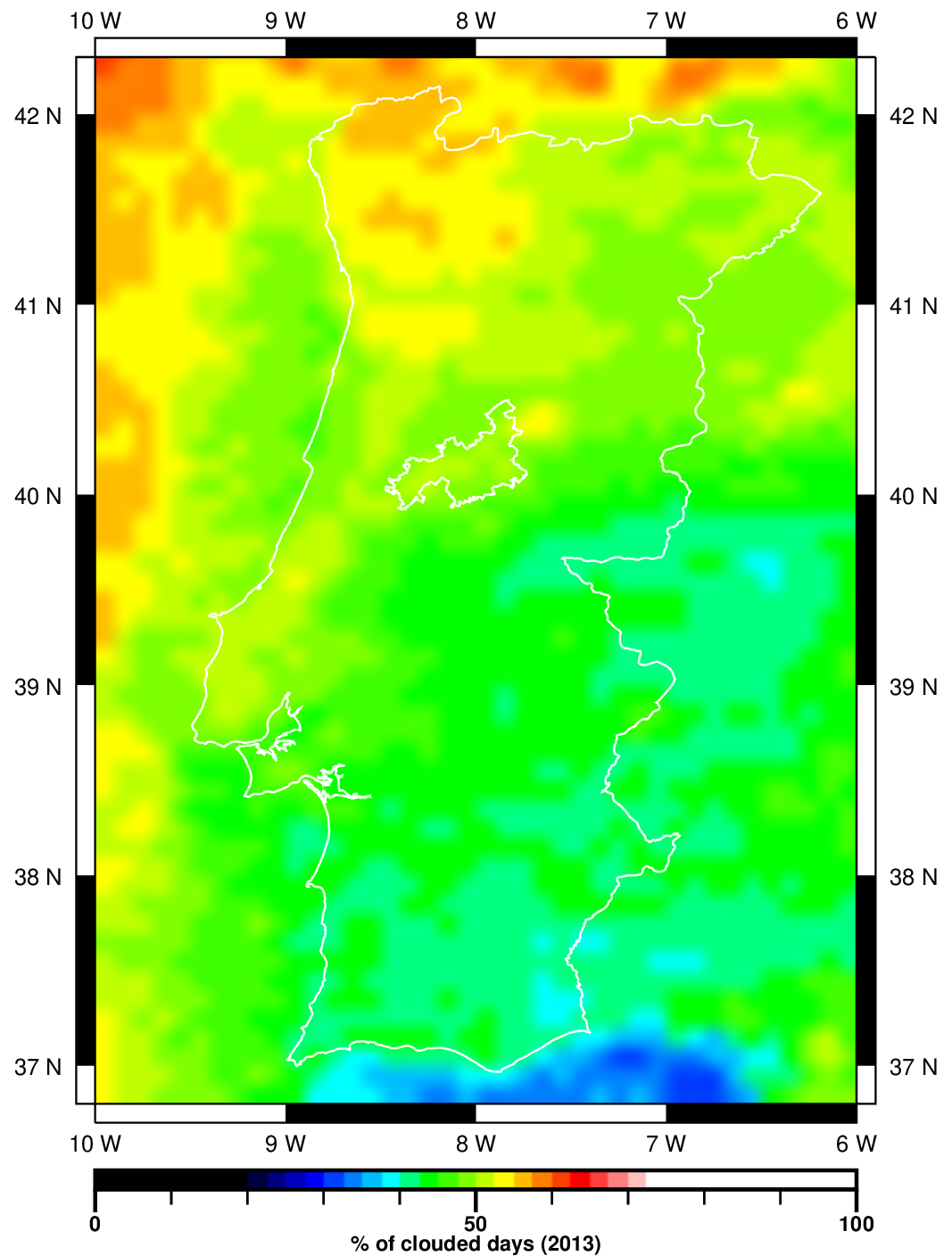}
\includegraphics[width=.3\textwidth]{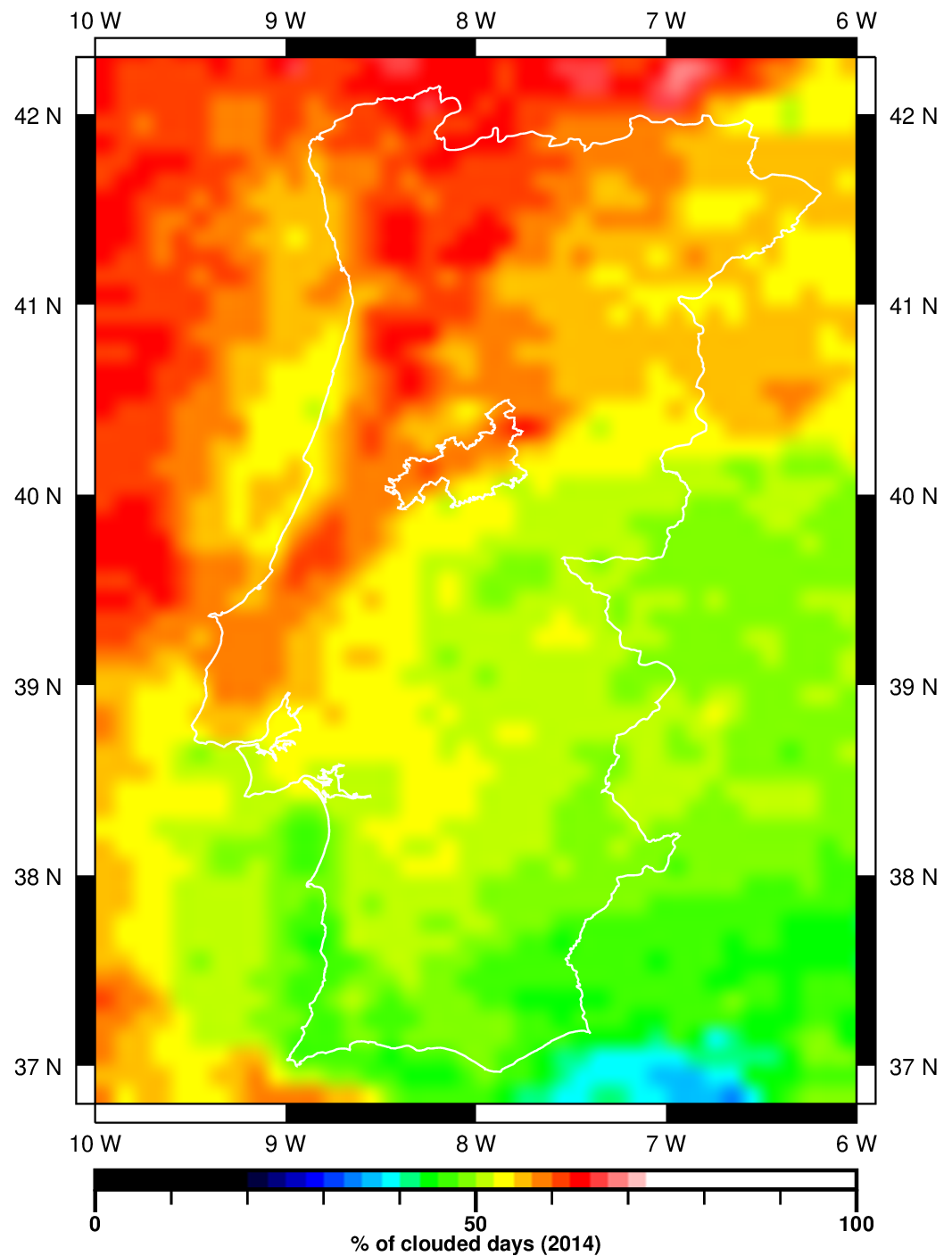}
\end{center}

\begin{center}
\includegraphics[width=.3\textwidth]{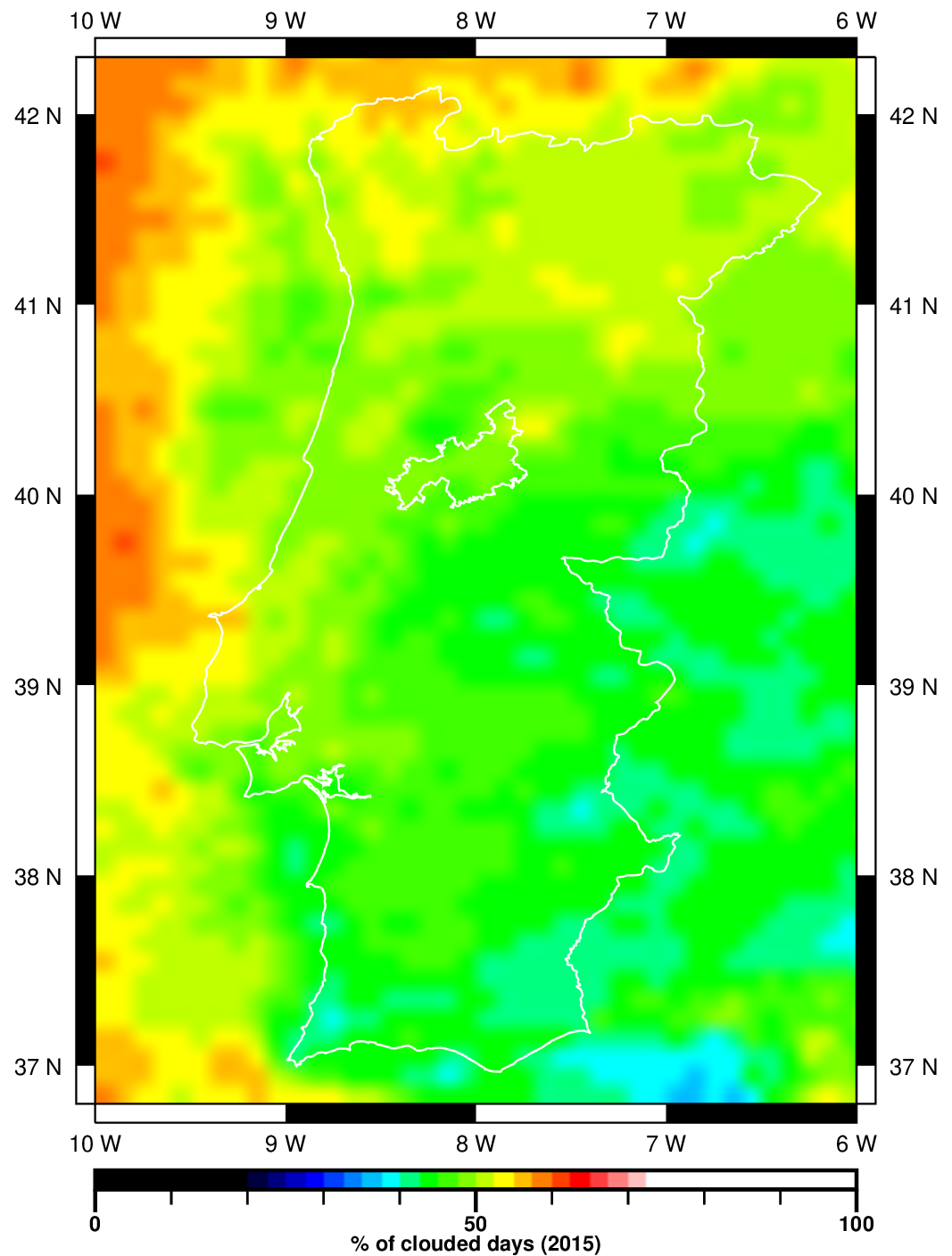}
\includegraphics[width=.3\textwidth]{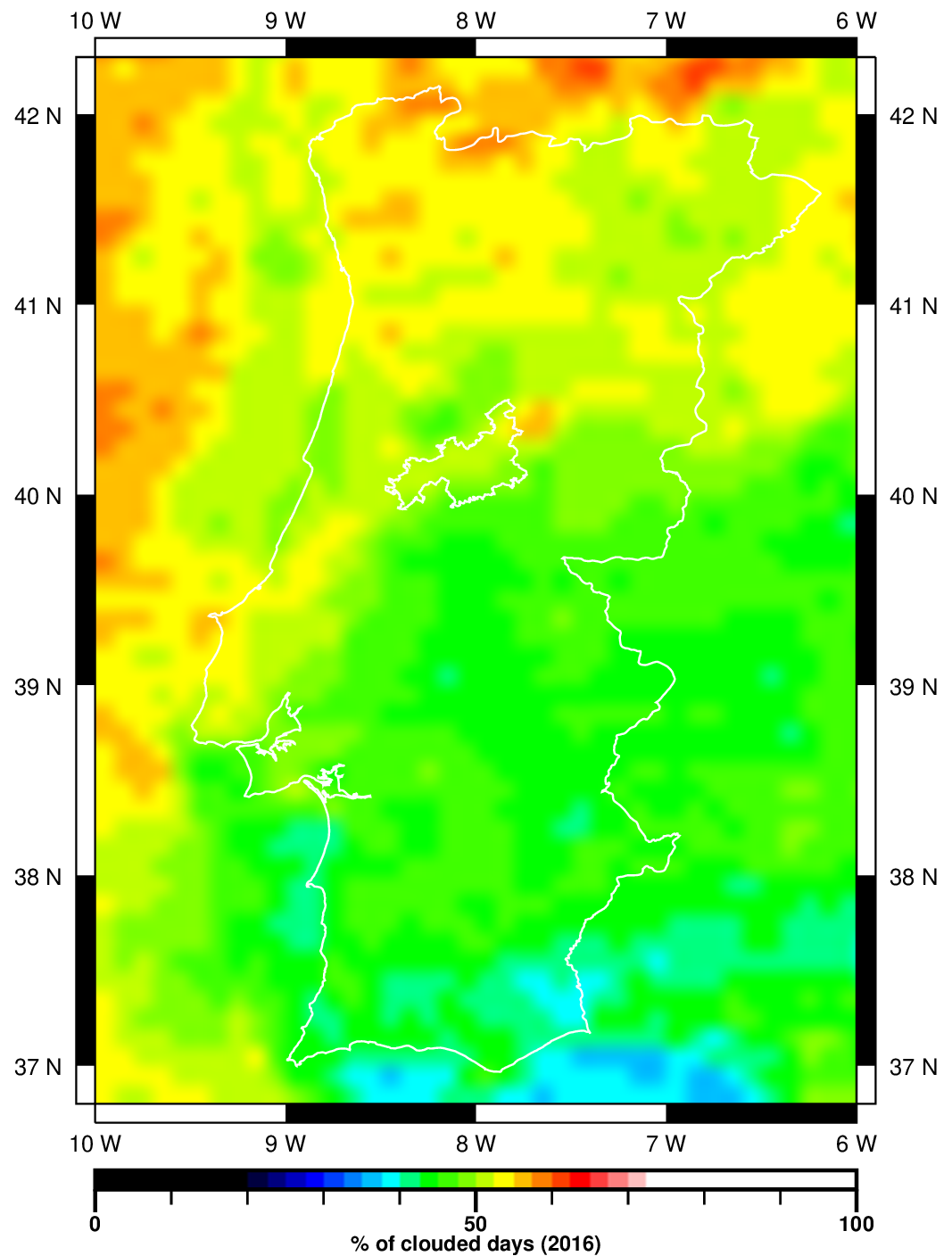}
\includegraphics[width=.3\textwidth]{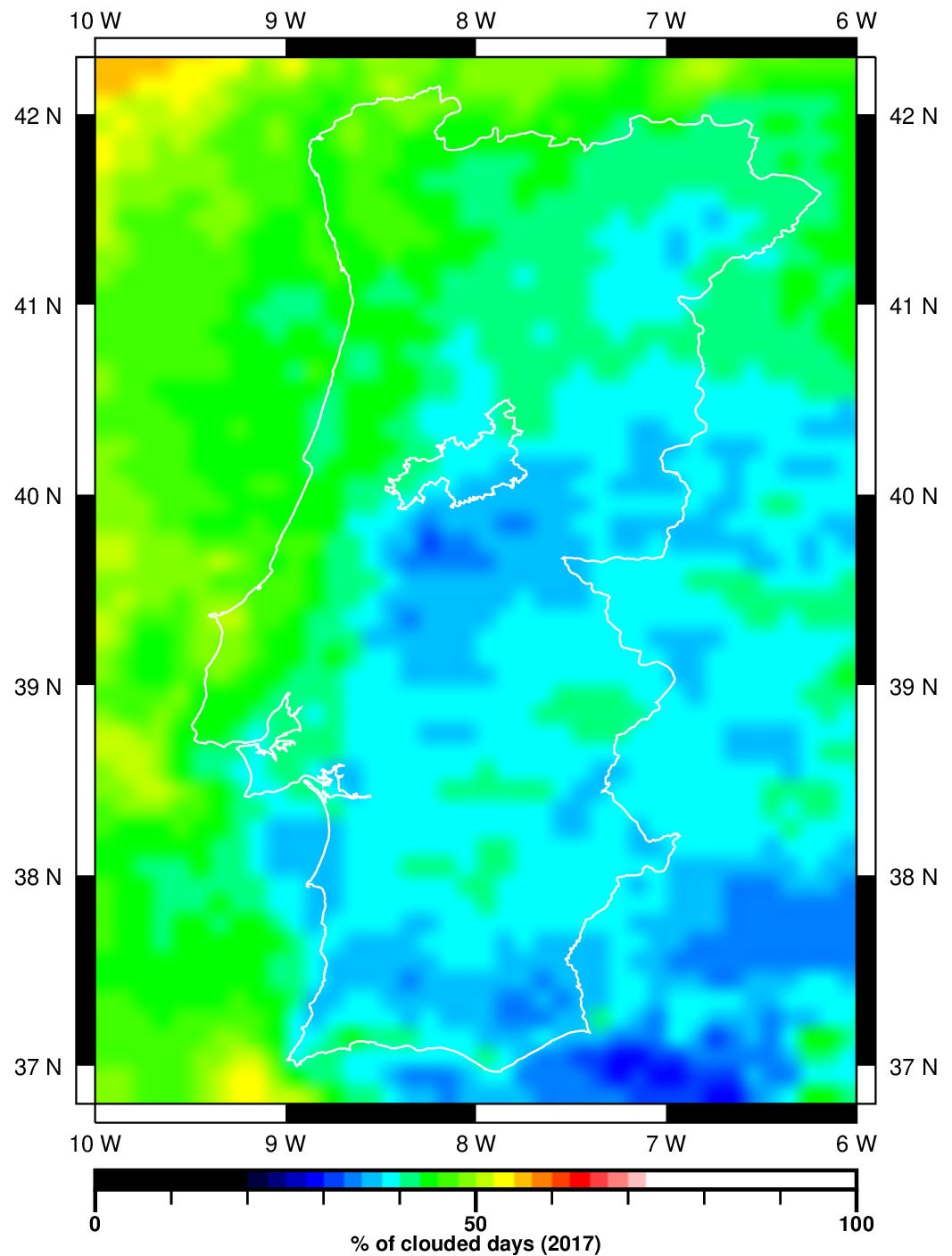}
\end{center}

\caption[MODIS/Aqua cloud cover maps for 2012-2017]{Maps of the number of the fraction of clouded days over Portugal for the years from 2012 to 2017, based on AQUA/MODIS data.}
\label{fg:modis}
\end{figure}

\begin{figure}
\begin{center}
\includegraphics[width=.85\textwidth]{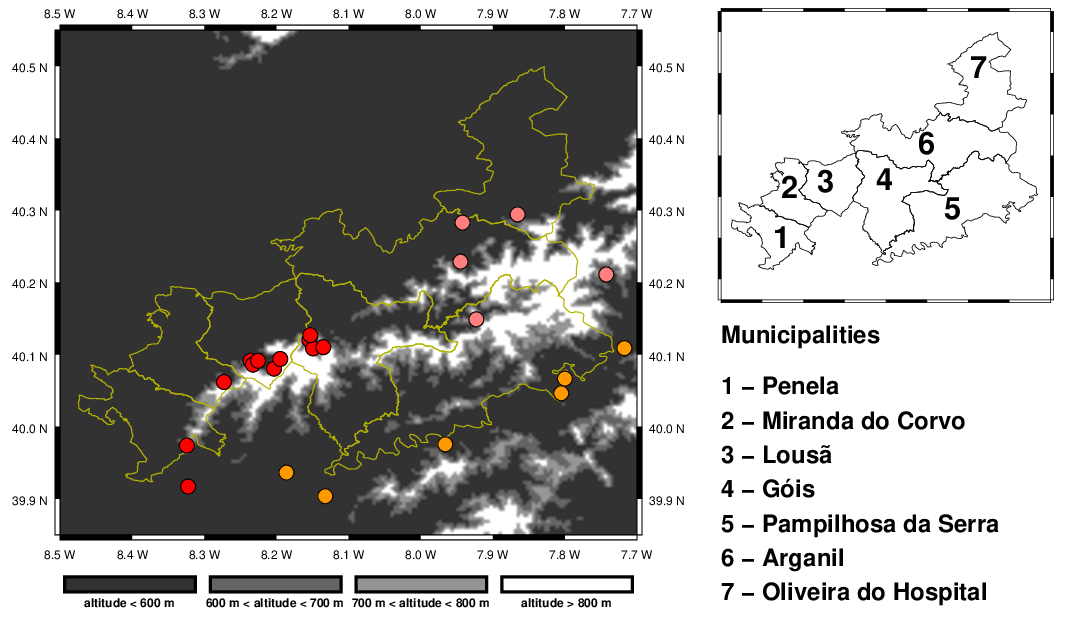}
\end{center}
\caption[Digital Elevation Model of the region]{Digital Elevation Model of the region.}
\label{fg:demcim}
\end{figure}

Figure~\ref{fg:modis} shows maps obtained from MODIS/AQUA data for the years from 2012 to 2017. Although at the scale of the whole country there are major variations, the reference councils show similar behavior in four of those six years. One of the two atypical years (2014) shows a much higher fraction of clouded days, while the other (2017) shows a much higher fraction of cloudless days. This picture is also seen when we look into the preceding years, and as such we infer that 2015 and 2016 represent fairly typical years, and that one every 3 to 4 years will be substantially cloudier or much less cloudier than those two.

\section{Sky brightness}\label{sec:skybrightness}

Detailed maps for the sky brightness in the CIM-Coimbra region were obtained from nighttime images of the the Portuguese territory provided by the Defense Meteorological Satellite Program (DMSP). The DMSP team provides monthly georeferenced mosaics, obtained by processing VIIRS data from selected nights. Cloudy Observations are removed and moonlight effects are corrected. For the latitudes of CIM-Coimbra these composites are available for all months excluding the period from May to August. In addition to these products, the DMSP team also provides yearly maps, where transient lights are filtered out and cloud cover information is made available. At the time of this report there were only two such yearly maps available, one for the year 2015 and another for the year 2016. Satellite observations are important for dark sky studies since those allow a precise analysis of the whole region of interest. Although the relative calibration of the pixels in the image is reliable, an absolute calibration still needs to be performed with in situ measurements at specific locations, to check that true magnitudes in the V band can be assigned to each pixel in the image. 

\begin{table}
\caption{In situ measurements of sky brightness.}
\label{tb:bright}
\begin{center}\begin{tabular}{@{} l *4c @{}}
\toprule
 \multicolumn{1}{c}{Site}    &  Coordinates & \multicolumn{3}{c}{Sky magnitude at Zenith} \\
   &   & Dec 2016   & Dec 2017  & July-Aug 2018  \\\midrule
Cabeço da Urra & 40\textdegree 01’31”N, 07\textdegree 56’52”W & 21.10 & - & - \\
Pessegueiro &	40\textdegree 03’42”N, 08\textdegree 00’58”W &21.40 & - & - \\
Janeiro de Baixo& 40\textdegree 02’10”N, 07\textdegree 48’32”W&	21.40 & - & - \\
Casal da Lapa& 40\textdegree 06’06”N, 07\textdegree 50’39”W&	21.30 & - & - \\[1ex]

Portela de Unhais & 40\textdegree 03’11”N, 07\textdegree 49’44”W & 21.25 & 21.30 & - \\
Porto da Balsa & 40\textdegree 11’00”N, 07\textdegree 52’43”W & 21.30 & 21.20 & 21.35 \\
Fajão & 40\textdegree 08’42”N, 07\textdegree 55’03”W  &21.30 & 21.15 & 21.25 \\\bottomrule
\end{tabular}\end{center}
\end{table}

\section{In situ measurements for calibration}

In order to calibrate satellite data, monthly campaigns near the time of new moon were planned to measure brightness in selected dark locations. Measurements at each selected location were planned to be acquired at several places separated by a few hundred meters to check homogeneity of the brightness distribution on a scale of the order of the VIIRS pixel size (350 meters). The campaigns were planned since July 2017, but due to the devastating forest fires that took place in the region from June to October that year, the first field work campaign was only made by year’s end. By that time the situation in the inhabited areas was already close to normality but the satellite data and in situ measures for those years might not reflect the average conditions, so instead we decided to rely on results from a previous campaign which took place in the year 2016, when sky brightness measurements were taken with a Unihedron Sky Quality Meter (SQM) L device, during the new moon, January 2016. 

Some of the sites visited in 2016 were revisited during the new moon in December 2017 and on  August 2017 (here not yet new moon but moon was below the horizon). Results of the three campaigns are shown in Table~\ref{tb:bright}. At two of the sites, 
the sky brightness values were lower in December 2017, probably an effect of the October 2017 wild fires, but were again close to the 2016 values on August 2018. Table~\ref{tb:bright} shows only measured zenithal brightness, although measurements at 45 degrees pointing towards the East, West, North and South were also acquired. No large differences were seen, except in one case when planet Venus was in the field of view. The values seen in North and West directions were not significantly different from the zenithal values, but to the East and South the magnitude of the sky was on average lower by 0.15 mag, as expected since these directions point to the coastal regions with bigger settlements. 

\begin{figure}
\includegraphics[width=0.65\textwidth]{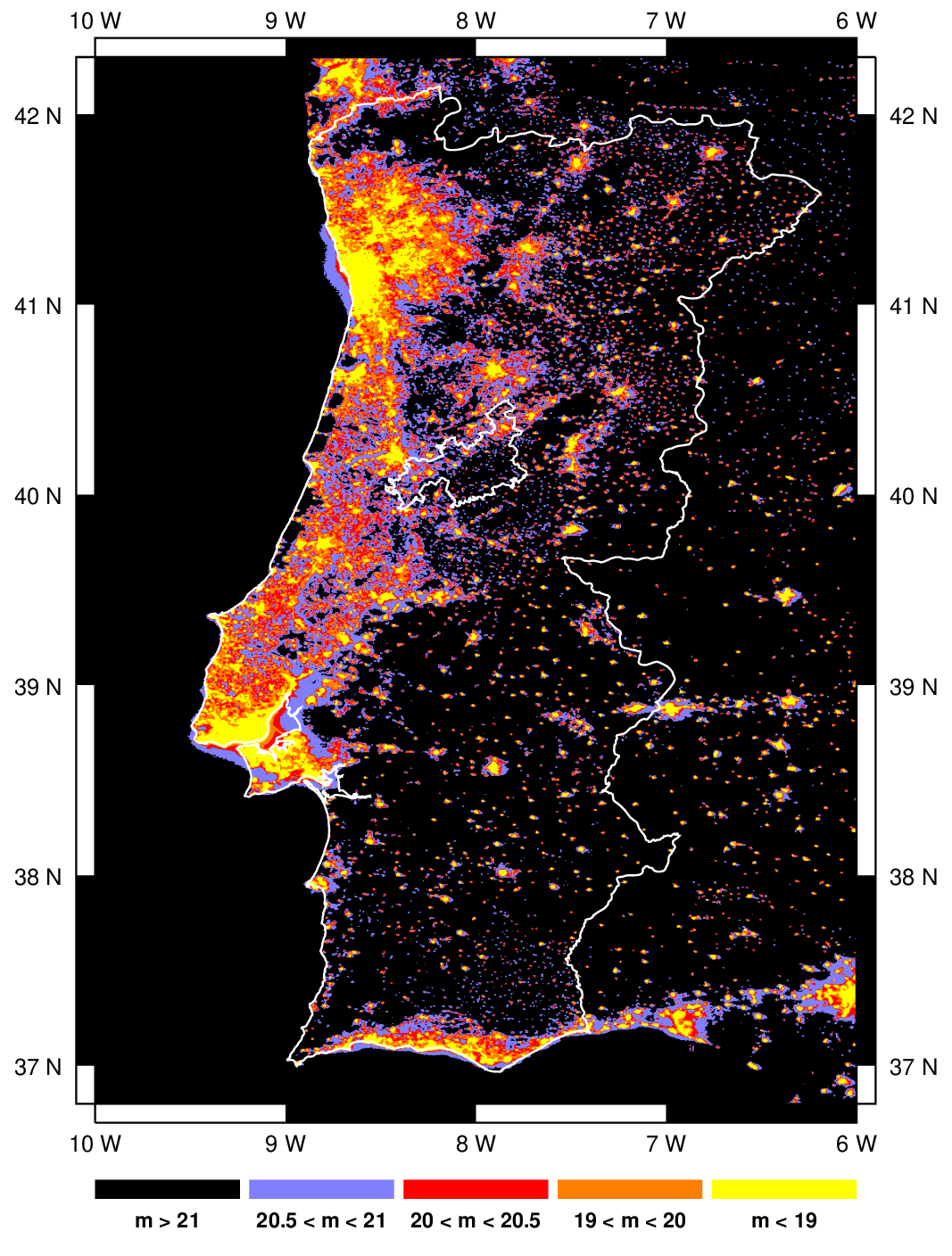}
\caption[Magnitude of the Sky over Portugal]{Map of the magnitude in the V band for the Portuguese mainland. The contours of the area defined by the seven councils of reference are superposed on the map.}
\label{fg:magpt}
\end{figure}

\subsection{Calibration of satellite data}

Using as reference for an absolute calibration the points in Table~\ref{tb:bright} brightness values in VIIRS data were converted into V band magnitudes both for the monthly map for December 2016 and for the yearly 2016 map. The yearly map was less noisy than the monthly map but, other than that, there were no major differences, so in what follows we will use the yearly map. We make no correction for possible differences in the VIIRS color response when compared with the equipment used for in situ measurements, since the departures that we expect at the very bright regions should not affect selection criteria. 

As shown in Figure~\ref{fg:magpt}, the Portuguese mainland presents a very bright littoral strip north of the city of Setúbal. Dark regions are seen in the South (except for the littoral strip of Algarve) and in the Interior Center regions. The CIM-Coimbra reference councils sit at the boundary between the bright littoral and the dark interior regions.

\begin{figure}
\includegraphics[width=\textwidth]{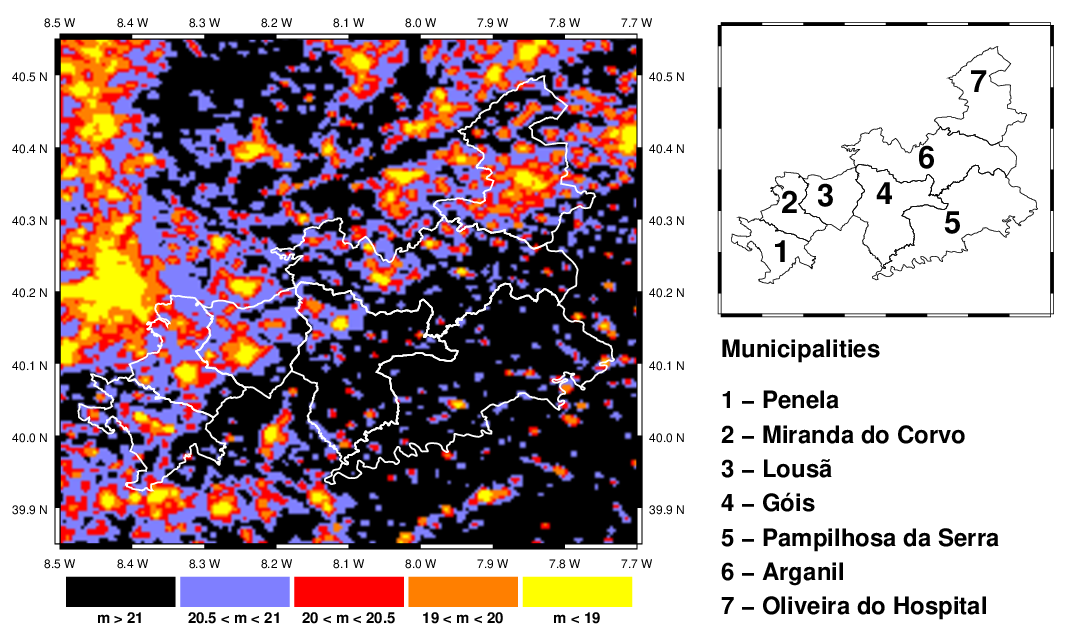}
\caption[Magnitude of sky over the region of interest]{Brightness map for the region of interest, with the contours of the seven reference councils shown. Only the black areas fulfill the criteria for certification.}
\label{fg:magcim}
\end{figure}

Figure~\ref{fg:magcim} provides a very detailed view (pixel size of 350 meters) over the CIM-Coimbra reference municipalities. Pampilhosa da Serra is mostly a dark sky municipality, while the other municipalities present extensive bright regions with inferred sky brightness below the 21 mag threshold.

\begin{figure}
\includegraphics[width=\textwidth]{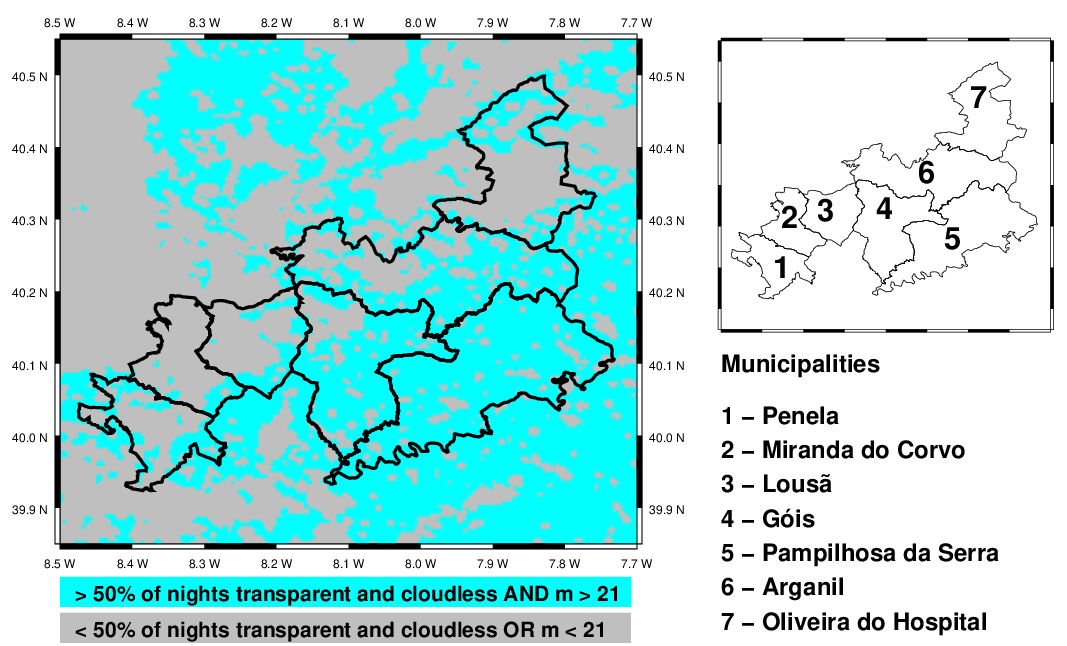}
\caption[Preliminary dark sky certifiable region]{Map of the region fulfilling simultaneously the sky brightness, transparency and cloud cover requirements for more than 50\% of the nights (in cyan).}
\label{fg:nuvmag}
\end{figure}

\section{Preliminary Selected regions}

Combining the analysis of brightness with the analysis of cloud cover and transparency discussed in previous sections we can now define the area respecting simultaneously the three dark sky certification criteria. Figure~\ref{fg:nuvmag} shows the contours of our reference councils in CIM-Coimbra over the region fulfilling simultaneously the sky brightness, transparency and cloud cover requirements for more than 50\% of the nights.
\begin{figure}
\includegraphics[width=\textwidth]{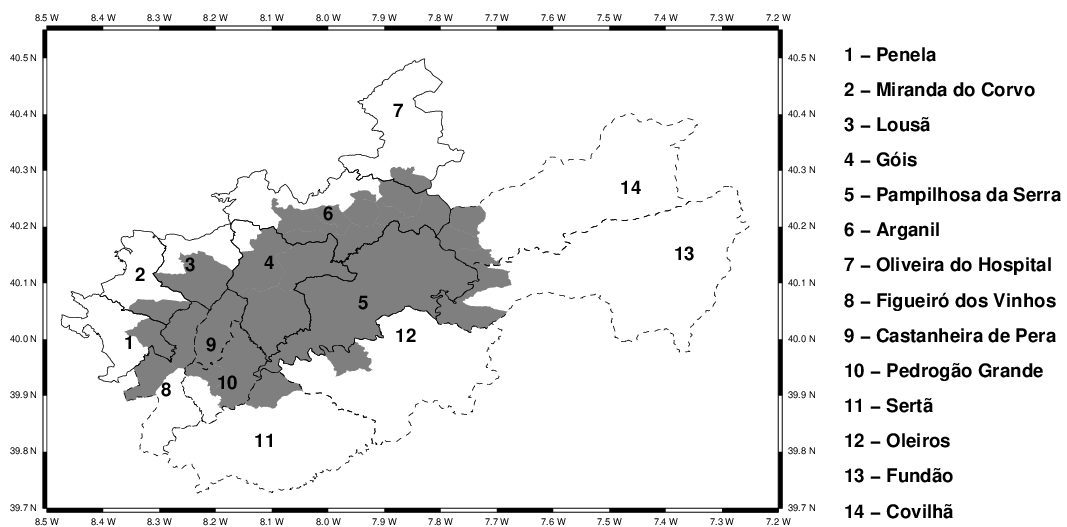}
\caption[Selected area]{Gray areas : selected areas for Dark Sky Starlight Tourist Destination.}
\label{fg:foracim}
\end{figure}

\begin{figure}[b]
\includegraphics[width=\textwidth]{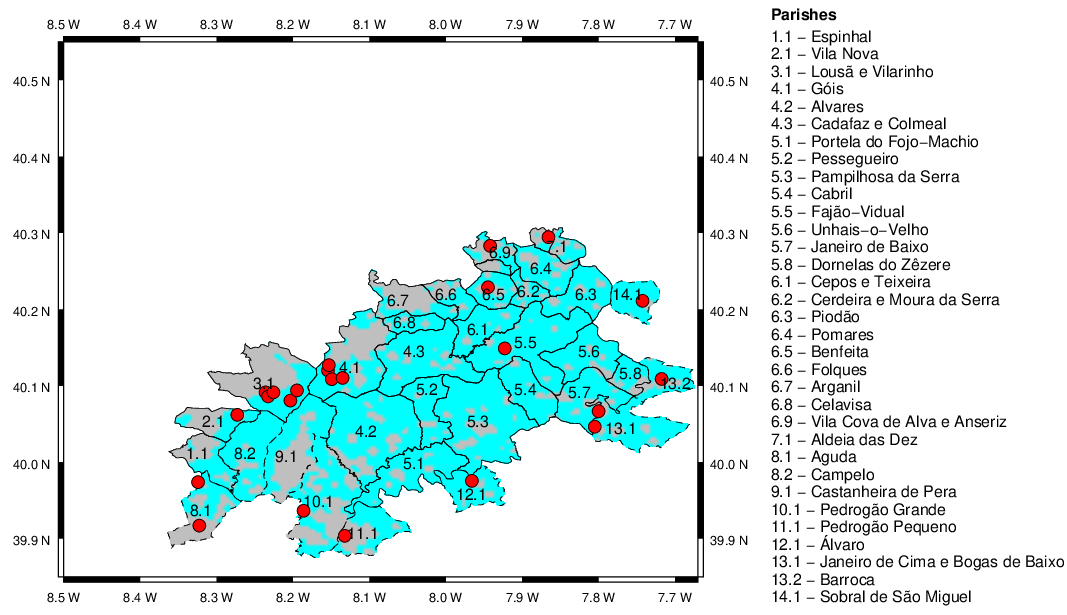}
\caption[Parishes in the contact area]{List of the 33 parishes in the contact region, and their contours superposed on the provisional certification map (in cyan). The red dots mark the locations of 23  Towns From Schist}
\label{fg:xistomais}
\end{figure}

We restricted our reference areas to the areas of the locations that present a sizable fraction of a potentially certifiable area, in dark blue. Even those areas that do not fulfill the seeing requirements, they might still be important as a buffer areas that needs to be protected. We define thus the area defined by the ensemble of those parishes as the "contact" area. Besides the parishes within the reference CIM-Coimbra municipalities we include parishes outside CIM-Coimbra that sit within the Serra da Lousã mountain range or include in their territory one or more Towns From Schist. The contact region includes thus 33 parishes. Figure~\ref{fg:foracim} shows the extent of this contact region, and the municipalities involved. We define the  provisional certification map as the area fulfilling simultaneously the sky brightness, transparency and cloud cover requirements for more than 50\% of the nights restricted to those 33 parishes. Figure~\ref{fg:xistomais} shows a list of the 33 parishes in the contact region, and their contours superposed on the provisional certification map. Figure~\ref{fg:xistomais} includes also the locations of 23 locations given in Table~\ref{tb:xisto}; they surround the provisional certification map.

\section{Seeing Measurements }

Standard atmospheric seeing measurement procedures rely on pointing to a region on the sky where one could image two stars of similar brightness, then track those two stars using a motorized mount and remove the slow frequency noise component of the seeing. Our setup included a reflector telescope with a 20 centimeter mirror, coupled to a camera capable of 30 images per second with integration times per image that could be as low as 1 ms. We tested the setup at the  Astronomical Observatory from Porto University where seeing at ground level is measured to be typically in the 3-7 arcsecond range prior to field work. The seeing computed at ground level was similar to the one we obtained on top of our Observatory tower, 6 meters above ground.

The planned field work was restricted to only a few good weeks in summer of July-August 2018 since 2017 year experienced devastating summer wildfires and 2018 turned to be clouded most of the spring and summer.


Since seeing has to be measured at ground level in many locations of a region with a difficult orography, we opted for a different approach. The technique we exploited was not to follow a pre-chosen star, but to point the telescope to a fixed sky position close to zenith, and then image a very bright star as it crossed the telescope field of view. The total seeing was then determined via linear least squares method to the centroid position of the star in each image. However, this method may be affected by vibrations caused by strong wind. Nevertheless, it is a much more practical method for field work since it does not involve a motorized mount, an advantage while traveling through the complicated orography and uneven terrain of the selected locations.  We refer the setup is also a low cost solution (< €900) enabling site seeing measurements for dark sky surveys by teams in regions where equipment acquisition is difficult or expensive. After our first field campaign, we gladly confirmed a large telescope is not required.  

In subsequent field work we used a simplified assembly consisting of a non-motorized mount and a refractor telescope with a 8 cm aperture and 0.6 meter focal length. To ensure that the pixel size in the images was such that the point spread function of the source had a diameter of at least 4-5 pixels, we placed a Barlow lens (2x) before the camera. With the typical image noise and background at the sites where we made observations, and for the stars that were our targets (Vega, Altair and other bright stars) we expected to get centroid positions with accuracy better than 0.05 arcsec. The equipment setup used in field work campaigns was the following:
\begin{itemize}
\item Orion 80 mm ED Refractor
\item Celestron OMNI 2x Barlow Lens
\item Orion StarShoot USB  Eyepiece II
\end{itemize}

\subsection{Field campaigns}

During this first field trip we executed our first seeing determination on a dark ($m_v$=21.2) open site (40.14505\textdegree N, 7.91761\textdegree W), about 750 meters from Fajão in Pampilhosa da Serra municipality. The derived seeing value was 1.42$\pm 0.02$ arcsec, much better than the threshold for certification (3 arcsec). 

\begin{figure}
\includegraphics[width=.9\textwidth]{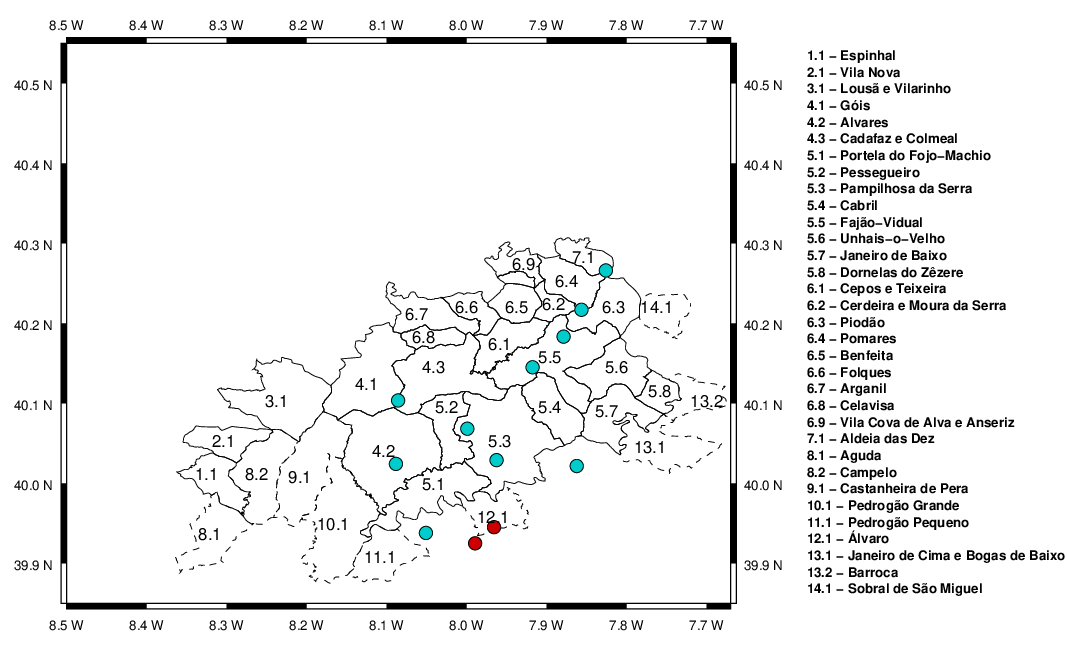}
\caption{Locations where the seeing was measured, superposed on a map. 
}
\label{fg:seeloc}
\end{figure}

We undertook ground measurements in 4 municipalities in CIM-coimbra: Pampilhosa da Serra, Góis, Arganil, and Oliveira do Hospital. We made also some measurements in the municipality of Oleiros, that, although not belonging to CIM-Coimbra, has a border with Pampilhosa da Serra. Sky brightnesses were measured at each and were between 21.2 and 21.4 magnitude. We emphasize that all sites that were chosen are exposed sites, that is, they are situated  close to main roads, in open areas, over shale soils with dry or burnt vegetation, or over dirty roads.  They reflect typical characteristics of dark sites and not of specific secluded spots and do sample an adequate range of altitudes (see Table~\ref{tb:seeing}).

\begin{table}
\caption{Results from the seeing field campaigns. All times are local times (UTC+1).}
\begin{center}
\begin{tabular}{@{} l *4c @{}}
\toprule
 \multicolumn{1}{c}{Site}    &  Coordinates & Seeing   & Altitude  & Date \& Time \\
     &   & (arcsec)   & (meters)  &   \\\midrule
 Aldeia Cimeira & 40\textdegree 01$^\prime$44$^{\prime\prime}$N, 07\textdegree 57$^\prime$45$^{\prime\prime}$W  & 2.28$\pm 0.06$ & 673 & 2018-07-28 23:24 \\ 
 Carvalho & 40\textdegree 04$^\prime$05$^{\prime\prime}$N, 07\textdegree 59$^\prime$57$^{\prime\prime}$W  & 2.43$\pm 0.05$ & 584 &  2018-07-30 00:41\\
Porto da Balsa & 40\textdegree 11$^\prime$00$^{\prime\prime}$N, 07\textdegree 52$^\prime$43$^{\prime\prime}$W  & 1.43$\pm 0.03$ & 840 & 2018-08-14 22:46\\
Piodão & 40\textdegree 13$^\prime$01$^{\prime\prime}$N, 07\textdegree 51$^\prime$23$^{\prime\prime}$W  & 1.47$\pm 0.12$ & 1142 &  2018-08-14 23:35\\
Aldeia das Dez & 40\textdegree 15$^\prime$57$^{\prime\prime}$N, 07\textdegree 49$^\prime$33$^{\prime\prime}$W  & 1.28$\pm 0.10$ & 1298 &  2018-08-15 00:18\\
Fajão & 40\textdegree 08$^\prime$42$^{\prime\prime}$N, 07\textdegree 55$^\prime$03$^{\prime\prime}$W  & 1.58$\pm 0.05$ & 862 &  2018-08-15 01:25\\
Álvares & 40\textdegree 01$^\prime$27$^{\prime\prime}$N, 08\textdegree 05$^\prime$17$^{\prime\prime}$W  & 3.52$\pm 0.10$ & 424 &  2018-08-15 23:21 \\
Cadafaz & 40\textdegree 06$^\prime$12$^{\prime\prime}$N, 08\textdegree 05$^\prime$08$^{\prime\prime}$W  & 1.40$\pm 0.11$ & 846 &  2018-08-16 00:17 \\
Madeirã & 39\textdegree 56$^\prime$16$^{\prime\prime}$N, 08\textdegree 03$^\prime$03$^{\prime\prime}$W  & 2.58$\pm 0.11$ & 601 &  2018-08-16 22:23 \\
Cambas & 40\textdegree 01$^\prime$17$^{\prime\prime}$N, 07\textdegree 51$^\prime$44$^{\prime\prime}$W  & 2.45$
\pm 0.06$ & 548 &  2018-08-17 00:06 \\\midrule
Sobral$^a$ & 39\textdegree 55$^\prime$30$^{\prime\prime}$N, 07\textdegree 59$^\prime$21$^{\prime\prime}$W  & 3.48$\pm 0.09$ & 903 &  2018-08-16 21:55 \\
Oleiros$^b$ & 39\textdegree 56$^\prime$42$^{\prime\prime}$N, 07\textdegree 57$^\prime$56$^{\prime\prime}$W  & 2.14$\pm 0.06$ & 937 &  2018-08-16 23:00 \\\bottomrule \\
 \end{tabular}

$^a$ seeing measured 230 meters away from wind turbine with rotating blades

$^b$ seeing measured 160 meters away from wind turbine with rotating blades
\end{center}
\label{tb:seeing}
\end{table}

\subsection{Seeing at higher altitudes}

From \ref{tb:seeing}, we can see two of the sites selected were situated at altitudes above 1000 meters: Piodão and Aldeia das Dez. 

These sites were selected for its altitude and because they were already protected areas. Both places had similar sky brightness values at about 21.2 mag, and wind blowing was weak. The seeing value computed for the Piodão site was 1.47$\pm 0.12$ arcsecond while for Aldeia das Dez site it was 1.28$\pm 0.10$ arcsecond. These values are much better than the threshold for certification (3 arcsecond). 
 
\begin{figure}
\begin{center}
\includegraphics[width=0.8\textwidth]{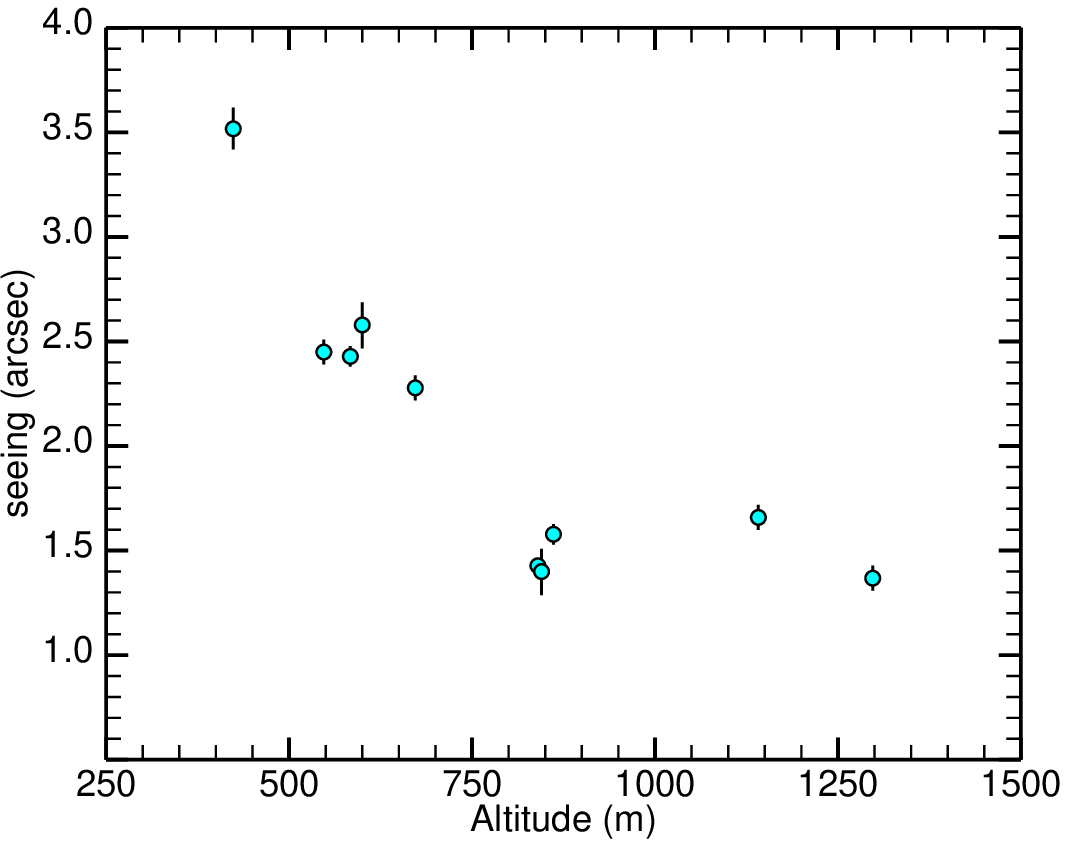}
\end{center}
\caption{Seeing versus altitude. Sites with altitude above 800 meters show seeing values around 1.5 arcsec. Below 800 meters seeing increases with decreasing height.}
\label{fg:seevsh}
\end{figure}

The best seeing values were obtained at the higher altitudes, in particular the lowest seeing value, 1.3 arcseconds, was obtained at 1300 meters near Aldeia da Dez, in the Oliveira do Hospital Municipality. We note that Fajão and Porto da Balsa present sky brightness consistent with the values obtained in December at the same sites, and that the seeing measured in Fajão in summer (1.58 arcsec) is close to the one seen in winter (1.42 arcsec).
The most obvious characteristic of our measurements is that seeing shows a strong dependence on altitude, as shown in Figure \ref{fg:seevsh}. 


\section{Summary and Outlook}

In this work we have described the development of a methodology integrating remote sensing techniques, open access space data, GIS environment tools and field testing for surveying light pollution and astronomical properties over a large area with irregular orography. Using satellite observations we were able to capture the average values and time variability of cloud cover and aerosol content over a large region in the Interior Center of Portugal. By combining night time satellite imagery with field testing we were also able to infer the sky brightness in magnitudes in the selected area of interest. The data covered the period 2012-2017 and established the methodology for regular light pollution and sky brightness studies in the region. The devised field testing is itself a low cost solution that can be used by teams in developing countries to easily assess dark sky potential prior to the development of local observatories and astrotourism activities. 

These techniques paved the way to a successful Dark Sky Tourist Destination award to the Portuguese "Aldeias do Xisto" (Schist Villages Regional Network) by the Starlight Foundation,  endorsed by UNESCO and UNWTO, in 2018 and renewed by 2023. This region has a considerable environmental interest with several NATURA 2000 areas, home to a high biodiversity and possessing some publicly protected relics of original "laurisilva" forest. This sparsely populated region has an almost absent industrial activity and therefore presents a strong potential for Dark Sky protection, thus enriching its environmental protection status. The procedures developed have been instrumental to study the nigh sky as a key environmental parameter and make Aldeias do Xisto  a case study of integrated environmental protection that includes light pollution control. 

\begin{figure}
\begin{center}
\includegraphics[width=0.9\textwidth]{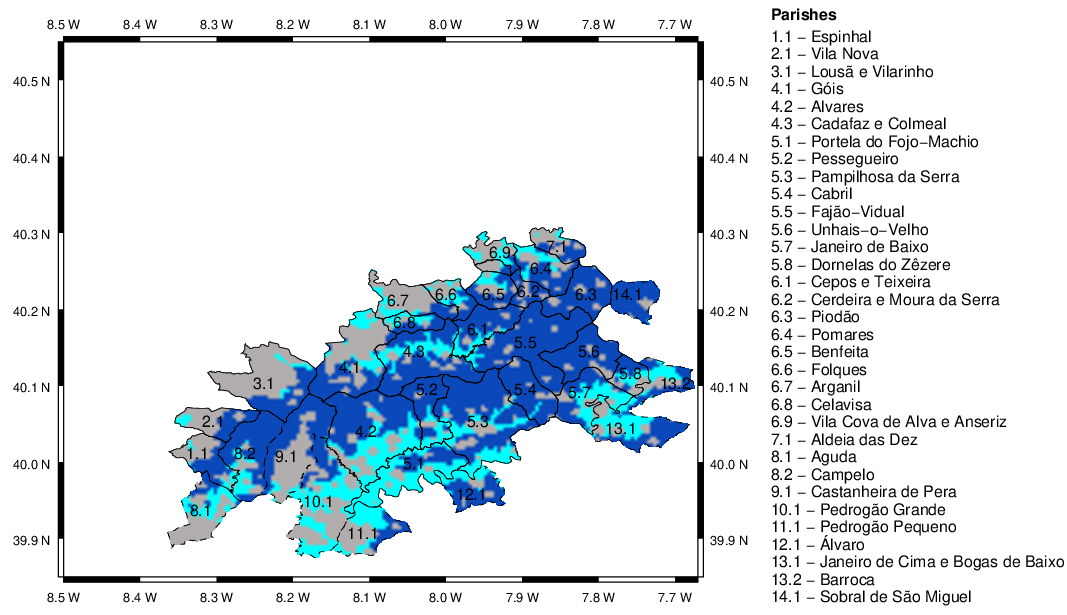}
\end{center}
\caption[Enhanced preliminary certification area]{Map showing in cyan the area where only the darkness, transparency and cloudiness thresholds are met and 
in blue an area where the seeing threshold is expected to be met also (corresponds to altitudes above 500m).}
\label{fg:certquin}
\end{figure}

The analysis revealed that an area corresponding to most of the Lousã and Açor Mountains at altitudes above 500 meters fulfill the criteria for Starlight certification regarding cloud cover, atmospheric transparency and sky brightness. The seeing measurements obtained in field campaigns confirmed the darkest locations are in the mountain ranges. This allowed us to propose an enhanced preliminary dark sky Starlight certification area that is shown in blue in Figure~\ref{fg:certquin}.

We note that the higher altitude of this mountainous region is the relevant factor making this region different from others Dark Sky destinations in Portugal that already exist in Portugal. Of particular interest is the region shown in Figure~\ref{fg:certquin} in a dark blue color. It marks the extent of the mountain ranges that are above 800 meters and with average seeing values measured around 1.5 arcsec.

Since the surveyed area is relatively large and with very complex orography, routine field testing in this region with specialized equipment proved to be both both time consuming and costly. Hence, ongoing automated monitoring and the use of GIS techniques with available satellite data are recommended as the standard method for local and continuous sky background monitoring. 

Over the last decade, forest wild fires were quite frequent in the region and the sky quality varied significantly during summer time with poor atmospheric conditions due to smoke and aerosols. Continuous monitoring of this mountainous region with adequate field testing recommends the development and use of regional climate models and the installation of strategically deployed automated stations to obtain long term time series of seeing and weather conditions with finer geographic granularity. This approach was developed for instance in the Canary Islands \citep{2014MNRAS.440.1964G} and Antarctica \citep{2016ThApC.125..841F}. 

\section*{Acknowledgements}
This work was supported through a research delivery contract funded by CIM-Coimbra, {\it “AD 21/2017 – Serviços técnicos especializados para a monitorizacão astronómica do céu na Comunidade Inter-Municipal da Região de Coimbra} to FCUP. We acknowledge support from ENGAGE SKA (POCI-01-0145-FEDER-022217), funded by COMPETE 2020 and FCT, Portugal. DM and BM acknowledge Project Centro de Investigação em Ciências Geo-Espaciais, reference UIDB/00190/2020, funded by COMPETE 2020 and FCT, Portugal. AKAJ and TV acknowledge through CFisUC  from FCT - Funda\c{c}\~ao para a Ci\^encia e a Tecnologia, I.P., Portugal, in the framework of the project UID/04564/2025, with DOI identifier 10.54499/UID/04564/2025. Partially funded by the Foundation for Science and Technology (FCT), I.P., through national funds, and by the European Regional Development Fund (ERDF) through the Innovation and Digital Transition Thematic Program (COMPETE 2030), of Portugal 2030, and by the European Union - Operation No. 14981 - COMPETE2030-FEDER-00860300 SPACE.








\bibliographystyle{elsarticle-harv}

\end{document}